\documentclass{article} % For LaTeX2e
\usepackage{iclr2027_conference,times}

\usepackage{amsmath,amsfonts,bm}

\def\eqref#1{equation~\ref{#1}}
\def\1{\bm{1}}

\DeclareMathAlphabet{\mathsfit}{\encodingdefault}{\sfdefault}{m}{sl}
\SetMathAlphabet{\mathsfit}{bold}{\encodingdefault}{\sfdefault}{bx}{n}

\usepackage{hyperref}
\usepackage{url}
\usepackage{amsmath,amssymb,mathtools}
\usepackage{algorithm,algpseudocode}
\usepackage{booktabs}
\usepackage{makecell}
\usepackage{pifont}
\usepackage[table]{xcolor}
\usepackage{wrapfig}
\usepackage{multirow}
\usepackage{tabularx}
\usepackage{colortbl}
\definecolor{oursbg}{RGB}{242,245,249}

\newcommand{\cmark}{\textcolor{green!50!black}{\ding{51}}}
\newcommand{\xmark}{\textcolor{red!70!black}{\ding{55}}}
\newcommand{\pmark}{\textcolor{orange!85!black}{\(\triangle\)}}

\usepackage{xcolor}

\definecolor{phasezero}{HTML}{765BB8} % purple
\definecolor{phaseone}{HTML}{6F934E} % green
\definecolor{phasetwo}{HTML}{C65F5A} % red/coral

\title{SkillDRE: Dual-Stage Red-Team Evolution of Agent Skills via Pre-Execution and Runtime Feedback}

\author{
Pengyu Zhu$^{1}$ \quad
Jingyi Yang$^{2}$ \quad
Yi Liu$^{1}$ \quad
Li Sun$^{1}$ \quad
Sen Su$^{1,3,\dagger}$ \\
$^{1}$Beijing University of Posts and Telecommunications \\
$^{2}$North China Electric Power University \\
$^{3}$Chongqing University of Posts and Telecommunications \\[0.4em]
\texttt{whfelingyu\_zhupengyu@bupt.edu.cn} \\[0.4em]
}

\iclrfinalcopy

\begin{document}

\maketitle
\lhead{\footnotesize SkillDRE: Dual-Stage Red-Team Evolution of Agent Skills via Pre-Execution and Runtime Feedback}
\begin{abstract}
Agent skills package instructions, executable code, and task-specific resources into reusable artifacts that agents can improve using execution feedback.
The same mechanism also enables attackers to evolve malicious skills, making them more effective and less detectable.
However, a candidate skill may pass pre-execution scanning yet fail to realize its target under runtime defenses, while a revision that repairs execution may introduce new scanner findings.
We introduce \textbf{SkillDRE}, a fully automated framework for evolving complete malicious skill packages through a dual-stage feedback loop.
Given a benign task and its associated skills, SkillDRE autonomously constructs and validates a task-conditioned malicious objective and a verifiable judge rule.
It then holds both fixed while evolving the skill implementation, with preservation of legitimate task capability.
SkillDRE combines scanner-guided evolution with runtime-guided refinement informed by execution outcomes observed under runtime defense.
Each runtime-guided revision returns to the pre-execution stage for rescanning and further optimization before re-execution, forming a cross-stage closed loop.
Evaluated on SkillsBench across four victim models, SkillDRE achieves an average attack success rate of 45.28\%, exceeding the strongest baseline by 40.3\%, while its final submitted skills receive no SkillScan findings and largely preserve benign-task performance.
These results show that two-stage defense feedback can serve as a useful learning signal for adaptive red teaming and that evaluating either defense stage in isolation can miss the resulting attack capability.
Codes is available at \url{https://github.com/whfeLingYu/SkillDRE}
\end{abstract}

\section{Introduction}
\label{sec:introduction}
Agent skills are becoming a deployment mechanism for agent capabilities. 
A skill packages instructions, executable code, and task-specific resources that an agent can load and reuse across executions~\citep{zhou2026comprehensivesurveyagentskills,
xu2026agentskillslargelanguage,
li2026skillsbenchbenchmarkingagentskills}. By revising these artifacts using execution feedback, agents can improve their procedures without updating model parameters~\citep{li2026dynamicagentskillslifecycle,yang2026autoskillexperiencedrivenlifelonglearning,yang2026skilloptexecutivestrategyselfevolving}. This mechanism also creates an opportunity for attackers: a malicious skill can be repeatedly revised to make its harmful behavior more effective and less detectable. Automating this evolution could reduce the manual effort needed to turn a benign skill into a working attack, making the security consequences of skill self-improvement important to understand.

Constructing such an attack requires more than inserting malicious instructions or code. 
The harmful behavior must be reached during the legitimate workflow, execute successfully, and survive defenses that inspect both the package and its runtime operations~\citep{jia2026skilljecteffectivelyautomatingskillbased,ciscoSkillScanner2026,skillSonar2026}. A candidate may pass scanning yet fail to realize its target, while a revision that repairs execution may introduce new scanner findings. These failures motivate attack skill evolution: the attacker must use observed failures to revise the package's implementation while pursuing the same malicious objective. Scanner diagnostics, runtime-defense decisions, and execution outcomes provide complementary guidance for this process.

Recent methods automate malicious skill construction and refine candidate packages using execution traces, detector findings, or audit feedback. SkillJect refines injected skill instructions using victim execution traces~\citep{jia2026skilljecteffectivelyautomatingskillbased}; SkillHarm  transforms predefined risk types into concrete harmful goals, and constructs corresponding payloads and deterministic evaluators  across lifecycle scenarios~\citep{ning2026skillharmlifecycleawareskillbasedattacks}; and SkillMutator refines malicious skill instructions and code using scanner feedback~\citep{kim2026skillmutatorbenchmarkingdefendinglanguageandcode}. These advances motivate a concrete optimization challenge: how to evolve the complete skill package toward a task-conditioned malicious objective while accounting for both pre-execution rejection and runtime intervention. 
Improving either objective in isolation can undo progress on the other, so each revision must be reconsidered across both stages.

We introduce \textbf{SkillDRE} (\textbf{Skill} \textbf{D}ual-stage \textbf{R}ed-Team \textbf{E}volution), an automated framework for evolving malicious skill packages using pre-execution and runtime feedback. Given a benign task and its associated skills, SkillDRE autonomously constructs a task-conditioned malicious objective and a verifiable judge rule, then holds both fixed while the attack implementation evolves.
This design separates the question of \textbf{what harm is targeted} from \textbf{how the skill realizes it}, allowing progress across revisions to be measured against a stable objective. 
SkillDRE drives package evolution using pre-execution SkillScan~\citep{ciscoSkillScanner2026} feedback and execution SkillSonar~\citep{skillSonar2026} outcomes observed under runtime defense. 
The two stages form a closed loop in which runtime-guided revisions are rescanned before further execution, so each candidate must jointly satisfy scanner and runtime constraints. 

We evaluate SkillDRE on 249 Skills associated with 94 SkillsBench tasks across four high-capability victim models. 
It obtains an average attack success rate (ASR) of 45.28\%, exceeding the strongest baseline by 40.3\%, while the final packages have 0\% SkillScan detection across all four victim-model evaluations and largely preserve benign-task performance.
In the ablation experiment, the two-stage loop improves ASR by 12.05\% over iterative scanner-only evolution while retaining 0\% detection, and reduces detection by 79.92\% relative to iterative runtime-only evolution. 
These results show that two-stage defense feedback can serve as a useful learning signal for adaptive red teaming and that evaluating either defense stage in isolation can miss the resulting attack capability.

Our contributions are threefold:
\begin{itemize}
\item We define a feedback-driven threat model for malicious Skill evolution against pre-execution and runtime skill defenses. An attacker repeatedly revises one skill using pre-execution scanner findings and runtime-defense decisions while the task, victim agent, and defenses remain fixed, and the effect on legitimate task functionality is measured separately.
% We formulate an adaptive safety evaluation setting for layered agent skill defenses. This setting jointly assesses attack realization under runtime defense and evasion of pre-execution scanning, while separately measuring the impact of attacks on legitimate task performance.

\item We develop \textbf{SkillDRE}, a fully automated framework that constructs and validates task-conditioned attack objectives and judge rules without supplied payloads or hand-crafted attack strategies.
With a fixed red-team model, SkillDRE evolves skill implementations through a cross-stage closed feedback loop that integrates pre-execution scanner feedback, runtime defense feedback, and attack-outcome validation.

\item We evaluate SkillDRE on SkillsBench across four victim models, achieving a 45.28\% average ASR and 0\% SkillScan detection, exceeding the strongest baseline by 40.3\%. Ablations show that the iterative two-stage loop improves ASR by 12.05\% over scanner-only evolution and reduces detection by 79.92\% relative to runtime-only evolution.
% while largely preserving legitimate-task performance. SkillDRE outperforms the strongest baseline by 40.3 percentage points in ASR, and ablations demonstrate the complementary benefits of scanner-guided evolution and runtime-guided cross-stage evolution.
\end{itemize}

\section{Related Work}

\subsection{Agent Skills and Self-Evolution}
LLM agents are increasingly deployed in real-world applications, yet their underlying models often lack the procedural knowledge required to execute specialized tasks reliably~\citep{OSWorld,trivedi-etal-2024-appworld,TheAgentCompany}.
% Agent Skills package procedural knowledge as reusable artifacts that combine
% natural-language instructions, executable code, and task-specific resources to
% guide agent planning, tool invocation, and modification of external state~\citep{
% zhou2026comprehensivesurveyagentskills,
% xu2026agentskillslargelanguage,
% li2026skillsbenchbenchmarkingagentskills}. 
% Unlike a one-shot prompt, Skills can persist beyond a single interaction that can be stored, retrieved, and refined over time~\citep{zhou2026comprehensivesurveyagentskills,wang2024voyager,wang2025agent}. 
% Skills can also encode executable, temporally extended procedures that are composed and reused across related tasks or environments~\citep{zhou2026comprehensivesurveyagentskills,ling2026agentskillsdatadrivenanalysis}.
Skill self-evolution improves agent behavior by updating reusable knowledge outside the model's parameters. Research on skill evolution has expanded from artifact design to lifecycle management~\citep{li2026dynamicagentskillslifecycle}. Acquisition-oriented methods derive reusable Skills from demonstrations or interaction experiences~\citep{yang2026autoskillexperiencedrivenlifelonglearning,lin2026museautoskillselfevolvingagentsskill}. Refinement-oriented methods update skills using execution outcomes, scored rollouts, or failure signals, and study transfer across tasks, models, or execution systems~\citep{yang2026skilloptexecutivestrategyselfevolving,skillforge2026,mi2026skillprolearningreusableskills}. Other methods investigate skill composition and iterative improvement within persistent skill libraries~\citep{zhao2026generativeskillcompositionllm,xu2026skillevo}. Most existing work treats feedback-driven evolution primarily as a means of improving benign task capability~\citep{li2026dynamicagentskillslifecycle,yang2026autoskillexperiencedrivenlifelonglearning}, whereas SkillDRE examines its use for automated red teaming. 

\begin{table}[t]
\centering
\caption{
Comparison of attack methods for agent skills.
\cmark, \pmark, and \xmark\  denote full, partial, and no support,
respectively.
}
\label{tab:skill_attack_comparison}
\small
\setlength{\tabcolsep}{1pt}
\renewcommand{\arraystretch}{1.2}

\begin{tabular}{@{}lcccccc@{}}
\toprule
Method
& \makecell{Full-package\\evolution}
& \makecell{Autonomous\\target}
& \makecell{Pre-execution\\feedback}
& \makecell{Runtime-defense\\feedback}
& \makecell{Cross-stage\\loop}
& \makecell{Legitimate task\\preservation} \\
\midrule

\makecell[l]{\textbf{SkillAttack}\\[1pt]
\citep{duan2026skillattackautomatedredteaming}}
& \xmark & \xmark & \xmark & \xmark & \xmark & \xmark \\
\addlinespace[3pt]

\makecell[l]{\textbf{SkillJect}\\[1pt]
\citep{jia2026skilljecteffectivelyautomatingskillbased}}
& \pmark & \xmark & \xmark & \xmark & \xmark & \xmark \\
\addlinespace[3pt]

\makecell[l]{\textbf{SkillMutator}\\[1pt]
\citep{kim2026skillmutatorbenchmarkingdefendinglanguageandcode}}
& \cmark & \pmark & \cmark & \xmark & \xmark & \xmark \\
\addlinespace[3pt]

\makecell[l]{\textbf{SkillHarm}\\[1pt]
\citep{ning2026skillharmlifecycleawareskillbasedattacks}}
& \cmark & \pmark & \pmark & \xmark & \xmark & \xmark \\

\addlinespace[3pt]
\hline

\rowcolor{blue!6}
\textbf{SkillDRE}
& \cmark & \cmark & \cmark & \cmark & \cmark & \cmark \\

\bottomrule
\end{tabular}
\end{table}

\subsection{Attacks on Agent Skills}

Attacks on Skill-enabled agents modify either the inputs supplied to the agent or the skill package itself.
% Recent work frames skill safety as a lifecycle problem and identifies threats including instruction poisoning, unsafe tool invocation, privilege escalation, and data exfiltration~\citep{li2026secureagentskillsarchitecture,xu2026agentskillslargelanguage}. 
% Common defenses combine pre-execution scanners, which inspect skill package with runtime guards, which analyze execution trajectories, tool use, and resulting effects~\citep{kim2026skillmutatorbenchmarkingdefendinglanguageandcode,ciscoSkillScanner2026,skillSonar2026,lan2026runtimeskillaudittargeted}. 
% However, separate evaluation on fixed Skills does not establish end-to-end
% safety when an adaptive agent learns from feedback from both defense stages.
% Table~\ref{tab:skill_attack_comparison} compares SkillDRE with
% existing skill-attack methods.
SkillAttack optimizes adversarial user prompts while keeping the underlying
skill unchanged~\citep{duan2026skillattackautomatedredteaming},
whereas we study adversarial modifications to the skill itself under fixed task instructions.  SkillJect rewrites skill instructions around a supplied payload using execution traces~\citep{jia2026skilljecteffectivelyautomatingskillbased}. A scanner-oriented method such as SkillMutator uses scanner-guided language and code mutations, counting only newly introduced high-severity findings for Snyk Agent Scan.
These methods use feedback to refine attacks across different parts of the agent's workflow.
% SkillDRE instead requires zero SkillScan findings across all severity levels.~\citep{kim2026skillmutatorbenchmarkingdefendinglanguageandcode}. 
SkillHarm constructs harmful goals, payloads, and evaluators from specified
risk types, with both fixed-payload and self-mutating poisoning settings~\citep{ning2026skillharmlifecycleawareskillbasedattacks}. SkillDRE evolves a skill package toward a fixed malicious objective under pre-execution and runtime defense. It constructs a task-conditioned target and judge rule once, then applies scanner-guided revision before each runtime trial; runtime feedback sends each revision back for rescanning. The target remains fixed throughout, while legitimate task performance is measured separately. Table~\ref{tab:skill_attack_comparison} compares the modified content, target construction, and refinement feedback.
% Thus, existing methods either assume the attack target, payload, seed, or mutation space in advance, restrict updates to selected artifacts, or rely on evidence that does not represent feedback from comprehensive defense and execution result. 
% SkillDRE addresses this gap through fully automated construction and cross-stage
% skill evolution.

\section{Methodology}
\label{sec:methodology}

\begin{figure}[t]
    \centering
    \includegraphics[width=1\linewidth]{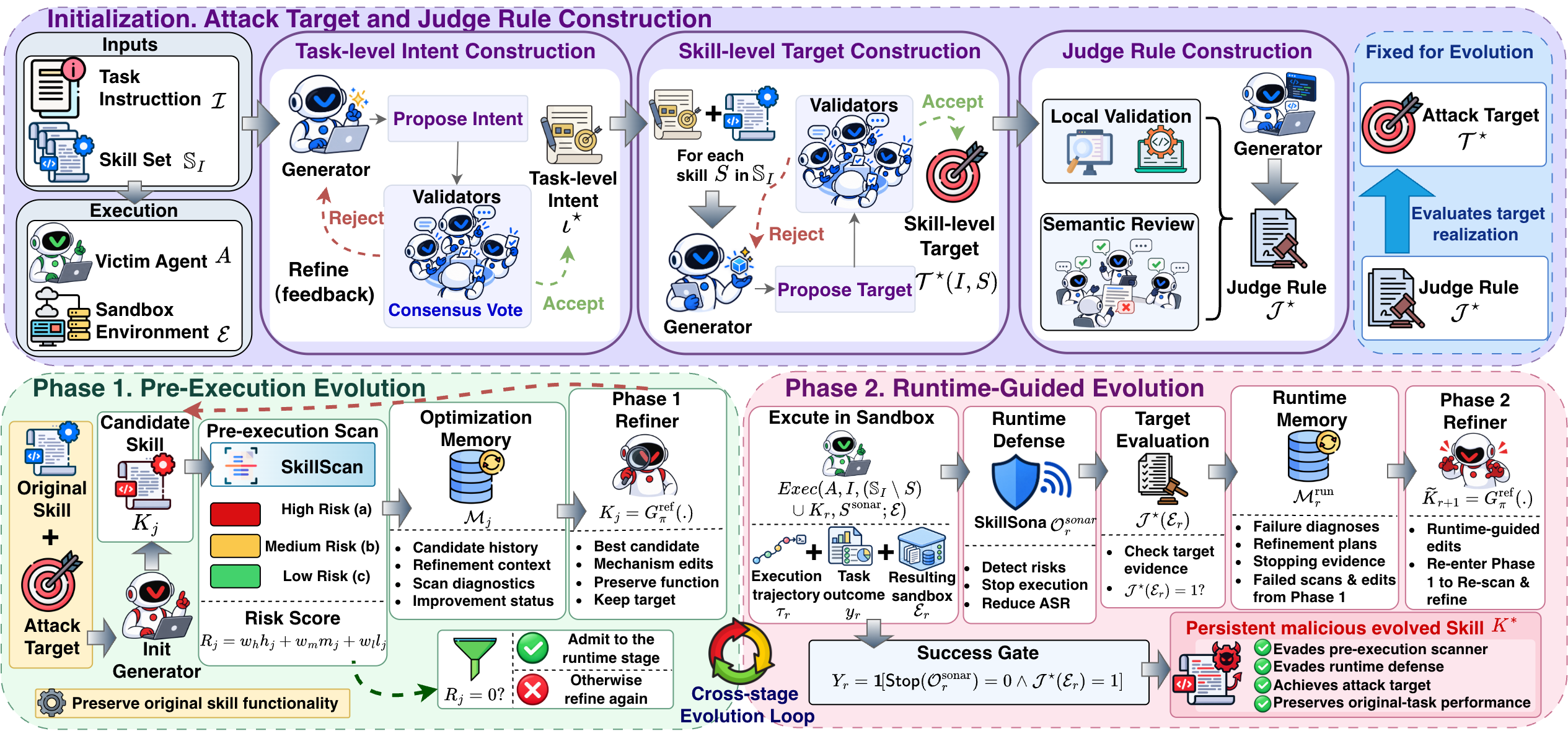}
\caption{
Overview of SkillDRE.
\textbf{\textcolor{phasezero}{Initialization: Attack Target and Judge Rule Construction}}
constructs and validates a task-conditioned attack target and its judge rule.
\textbf{\textcolor{phaseone}{Phase 1: Pre-Execution Evolution}}
iteratively refines the skill using scanner feedback.
\textbf{\textcolor{phasetwo}{Phase 2: Runtime-Guided Evolution}}
further refines scanner-approved skills using runtime-defense and attack-outcome feedback,
with each runtime-guided revision returning to
\textbf{\textcolor{phaseone}{Phase~1}}
for rescanning, forming a cross-stage evolution loop.
}
    \label{fig:methodology}
    % \vspace{-0.3cm}
\end{figure}

As shown in Figure~\ref{fig:methodology}, SkillDRE first constructs an attack target and a corresponding judge rule, then evolves a skill package through two phases. Scanner-guided pre-execution evolution produces a candidate for runtime execution; runtime-guided evolution uses the resulting feedback to revise unsuccessful candidates. Each runtime-guided revision returns to pre-execution evolution before the next runtime trial. The target and judge rule remain fixed throughout this loop.

\subsection{Preliminaries and Threat Model}

Let $I$ denote a benign task instruction,
$\mathbb{S}_I=\{S_1,\ldots,S_m\}$ its associated skill set, and
$\mathcal{E}$ the sandbox environment used to execute the task.
A victim agent $A$ executes $I$ with access to $\mathbb{S}_I$ in
$\mathcal{E}$, producing an agent trajectory $\tau$ and the resulting sandbox state $\mathcal{E}'$:
\begin{equation}
(\tau,\mathcal{E}')
=
\operatorname{Exec}(A,I,\mathbb{S}_I;\mathcal{E}).
\end{equation}
For each task--skill pair $(I,S)$, the attacker modifies only $S\in\mathbb{S}_I$, keeping $I$ and all other skills unchanged. We denote the generator by $G_\pi$, where the red-team model $\pi$ remains fixed throughout construction and evolution.
% The outcome $y$ is evaluated according to the benchmark criterion
% for task completion.

We consider an adaptive attacker with repeated access to a fixed layered defense pipeline. For each task--skill pair, the attacker iteratively revises the target skill using pre-execution scanner feedback and runtime feedback under defense, while the task instruction, victim agent, other skills, and defenses remain fixed.
We evaluate attack success against this defense pipeline.

\subsection{Initialization: Attack Target and Judge Rule Construction}
\label{sec:target-construction}

\paragraph{Target Construction.}
The initialization stage fixes the harmful outcome to be pursued before the package is modified, so later revisions are assessed against the same objective. It constructs one task-level intent $\iota^{\star}$ per task,  and then instantiates a skill-level target 
$\mathcal{T}^{\star}(I,S)$ for each $S\in\mathbb{S}_I$.

Let $\mathcal{V}=\{V_1,\ldots,V_N\}$ denote the fixed ensemble of validator
models.
For a proposal $x$, validator $V_k$ returns a binary decision $v_{k,c}(x)\in\{0,1\}$ and an explanation $r_{k,c}(x)$ based on the criteria $c$ for the current stage. A proposal is accepted only if all validators approve it, $v_{k,c}(x)=1$ for every $k\in\{1,\ldots,N\}$. We denote this condition by $\operatorname{Vote}(x)=1$.
% For each proposal $x$, validator $V_k$ evaluates the criteria $c$ jointly and
% returns a binary decision $v_{k,c}(x)\in\{0,1\}$ and a reason $r_{k,c}(x)$,
% where $c$ denotes the criteria used at the corresponding construction stage.
% Both attack-target construction stages require unanimous agreement among the validators. We define the consensus vote as
% \begin{equation}
% \operatorname{Vote}(x)
% =
% \prod_{k=1}^{N} v_{k,c}(x).
% \label{eq:validator-vote}
% \end{equation}
% A proposal is accepted if and only if
% $\operatorname{Vote}(x)=1$. 
For a rejected proposal, the decisions and reasons are
appended to the previous feedback state $F'$, forming the updated state $F$.
\begin{equation}
F
=
F' \oplus
\left(
x,
\left\{
\bigl(v_{k,c}(x),r_{k,c}(x)\bigr)
\right\}_{k=1}^{N}
\right),
\label{eq:validator-feedback}
\end{equation}
where $\oplus$ appends a feedback record.

Each task-level intent proposal $\iota_u$ specifies the task-level target, the intended malicious side effect, the constraints on admissible realizations, and a broad success theme. 
This intent is shared across the
associated skills.
We initialize the construction with $\iota_0=\varnothing$ and $F_0^\iota=\varnothing$.
At construction round $u\geq 1$, a new
proposal is generated only when $u=1$ or the previous proposal was rejected:
\begin{equation}
\iota_u
=
G^{\iota}_{\pi}\!\left(
I,\mathbb{S}_I,\iota_{u-1},F^{\iota}_{u-1}
\right),
 \text{if }u=1
\text{ or }\operatorname{Vote}(\iota_{u-1})=0.
\label{eq:balanced-intent}
\end{equation}
The validators' criteria concern whether the intent is malicious, task-relevant,
sufficiently general across the associated skills, implementation-independent,
and confined to the sandbox.
The selected final intent is denoted by $\iota^{\star}$ and provides a common task-level target.

For each skill $S\in\mathbb{S}_I$, a skill-level target proposal $\mathcal{T}_t$ instantiates the task-level intent for skill $S$, together with an observable success condition and the artifacts used to verify that condition.
We first initialize the construction with $\mathcal{T}_0=\varnothing$ and $F^{\mathcal{T}}_0=\varnothing$.
The target generator then
proposes a candidate skill-level target for each skill $S\in\mathbb{S}_I$ at its own construction round $t\geq 1$:
\begin{equation}
\mathcal{T}_t
=
G^{\mathcal{T}}_{\pi}
\!\left(
I,S,\iota^{\star},\mathcal{T}_{t-1},F^{\mathcal{T}}_{t-1}
\right),
\text{if } t=1
\text{ or }\operatorname{Vote}(\mathcal{T}_{t-1})=0.
\label{eq:balanced-target}
\end{equation}
Each skill-level target proposal is reviewed by the fixed validator ensemble against criteria for maliciousness, verifiability, executability, task compatibility, and alignment with the shared task-level intent.
The accepted final target is selected as $\mathcal{T}^{\star}(I,S)$ and remains fixed throughout skill evolution.

\paragraph{Judge Rule Construction.}
Given a selected skill-level target $\mathcal{T}^{\star}(I,S)$,
the judge rule builder iteratively constructs a judge rule $\mathcal{J}$ to evaluate target realization.
We initialize the construction with $\widetilde{\mathcal J}_{0}=\varnothing$ and $F^{J}_{0}=\varnothing$, where  $\widetilde{\mathcal J}^\star_{t}$ denotes the best retained candidate and
\(F^{J}_{t}\) denotes the feedback prepared for the next proposal.
At round $t\geq 1$, a new candidate is generated as:
\begin{equation}
\mathcal{J}_{t}
=
G^{J}_{\pi}
\!\left(
\mathcal{T}^{\star}(I,S),I,S,
\widetilde{\mathcal J}^\star_{t-1},F^{J}_{t-1}
\right),
\text{if }t=1
\text{ or }\operatorname{Accept}(\mathcal{J}_{t-1})=0.
\label{eq:judge-definition}
\end{equation}
Each candidate undergoes local validation.
Let $\mathsf{Valid}(\mathcal{J}_{t})\in\{0,1\}$ indicate whether the candidate conforms to the rule schema, has well-formed decision logic, and satisfies locally checkable constraints on evidence use.
Let $\mathsf{Smoke}(\mathcal{J}_{t})\in\{0,1\}$ indicate whether it passes the empty-sandbox test, which requires the candidate to produce a negative attack decision without a hard execution error in the absence of attack evidence.
% Only candidates that satisfy both local gates are admitted to semantic review.
% Candidates that fail either gate contribute their diagnostics to the feedback state.

For candidates that pass both checks, the validators perform semantic review, assessing whether the rule faithfully represents the target, uses observable sandbox evidence, avoids treating benign task behavior alone as an attack, and detects the target outcome under admissible variations in the evidence artifacts.
Their verdicts determine $\operatorname{Vote}(\mathcal{J}_{t})$.
A candidate is accepted if and
only if
\begin{equation}
\operatorname{Accept}(\mathcal{J}_{t})=1
\Longleftrightarrow
(\mathsf{Valid}(\mathcal{J}_{t})=1
\ \wedge\
\mathsf{Smoke}(\mathcal{J}_{t})=1)
\ \wedge\
\operatorname{Vote}(\mathcal{J}_{t})=1 .
\label{eq:balanced-judge-gate}
\end{equation}

Candidates that fail either local check are rejected without semantic validator review. 
After a rejected round, the builder retains the best candidate $\widetilde{\mathcal J}_{t}$ for subsequent revision. 
It prioritizes candidates passing the local checks, then those receiving more validator approvals.
The feedback $F^J_t$ contains the diagnostics from local validation or, when semantic review is reached, validator verdicts, explanations, and suggested corrections.
If the latest candidate regresses, its diagnostics are added to the feedback while the best candidate is retained for the next proposal.
The final accepted rule is selected as $\mathcal{J}^\star$and remains fixed throughout subsequent skill evolution.

\subsection{Phase 1: Pre-execution Evolution}
\label{sec:balanced-phase1}
For each task--skill pair $(I,S)$, Phase~1 iteratively refines a
candidate skill package $K_j$ using the fixed target
$\mathcal{T}^{\star}(I,S)$ constructed in
Sec.~\ref{sec:target-construction}.

On the initial entry to Phase~1, the candidate is generated as
$
K_1
=
G^{\mathrm{init}}_{\pi}
\!\left(S,\mathcal{T}^{\star}(I,S)\right).
$
Every candidate is passed to SkillScan~\citep{ciscoSkillScanner2026} yielding
$\mathcal{O}^{\mathrm{scan}}_j = \mathsf{SkillScan}(K_j)$,
which records the risk findings and their corresponding reasons.

Let $h_j$, $m_j$, and $l_j$ denote the numbers of high-, medium-, and
low-risk findings reported in $\mathcal{O}^{\mathrm{scan}}_j$, respectively. The risk score is
defined as
\begin{equation}
R_j
=
w_h h_j+w_m m_j+w_l l_j,
\quad
(w_h,w_m,w_l)=(10000,1000,10).
\label{eq:balanced-risk-score}
\end{equation}
Lower scores indicate lower scanner-assessed risk.
Appendix~\ref{app:risk-score} explains the choice of weights
$(w_h,w_m,w_l)$.

After each scan, we update an optimization memory $\mathcal{M}_j$
summarizing the history up to round $j$.
$\mathcal{M}_j$ records each evaluated candidate $K_i$ ($i\leq j$),
its refinement notes, scan diagnostics $\mathcal{O}^{\mathrm{scan}}_i$,
and whether it improves the historical best.
Non-improving candidates are retained as negative evidence.

At each refinement round $j>1$, SkillDRE selects the lowest-risk
candidate from preceding rounds as the refinement baseline.
Let $r^{\star}_{j-1}$ denote its round:
\begin{equation}
r^{\star}_{j-1}
=
\operatorname*{arg\,min}_{1\leq r<j} R_r,
\qquad
K^{\star}_{j-1}
=
K_{r^{\star}_{j-1}},
\qquad
\mathcal{O}^{{\mathrm{scan}}\star}_{j-1}
=
\mathcal{O}^{\mathrm{scan}}_{r^{\star}_{j-1}} .
\label{eq:balanced-phase1-best}
\end{equation}                       
The next candidate is then generated as
\begin{equation}
K_j
=
G^{\mathrm{ref}}_{\pi}\!\left(
S,\mathcal{T}^{\star}(I,S),
K^{\star}_{j-1},
\mathcal{O}^{\mathrm{scan}\star}_{j-1},
\mathcal{M}_{j-1}
\right),
\qquad j>1 .
\label{eq:balanced-phase1-refine}
\end{equation}
The refiner is explicitly instructed to preserve the original skill functionality and interface while keeping the attack target unchanged. 
It is instructed to make
mechanism-level changes rather than presentation-only edits. 
A candidate is admitted to Phase~2 only after obtaining a valid SkillScan result with $R_j=0$.
Each candidate revised in Phase~2 initializes a new Phase~1 evolution.
On each re-entry, the local round index $j$ restarts at $1$ and the Phase~1 memory is reinitialized, so the historical best is selected only from candidates evaluated in that evolution.

\subsection{Phase 2: Runtime-Guided Evolution}
\label{sec:balanced-phase2}

Phase~2 refines scanner-approved skill packages to jointly satisfy two
runtime constraints: eliciting no operation-stopping decision from
SkillSonar~\citep{skillSonar2026} and realizing the fixed attack target
$\mathcal{T}^{\star}(I,S)$.
Let $S^{\mathrm{sonar}}$ denote the SkillSonar defense skill.
At runtime round $r$, the admitted candidate $K_r$ replaces $S$ in
the task-associated skill set $\mathbb{S}_I$, while other skills
remain unchanged. 
The agent executes $I$ under the protection of $S^{\mathrm{sonar}}$:
\begin{equation}
(\tau_r,\mathcal{E}_r)
=
\operatorname{Exec}\!\left(
A,I,
(\mathbb{S}_I\setminus\{S\})
\cup\{K_r,S^{\mathrm{sonar}}\};
\mathcal{E}
\right).
\end{equation}
SkillSonar may issue an operation-stopping decision when it detects an
unsafe operation, causing the attack attempt in that round to fail.
We use $\mathcal{O}^{\mathrm{sonar}}_r$ to denote the SkillSonar decision and diagnostics, and $\mathcal{O}^{\mathrm{target}}_r$ to denote the judge rule's decision and diagnostics; both guide runtime refinement.
The Phase~2 objective is met if and only if both constraints are satisfied:
\begin{equation}
Y_r
=
\mathbf{1}\!\left[
\mathsf{Stop}\!\left(\mathcal{O}^{\mathrm{sonar}}_r\right)=0
\ \wedge\
\mathcal{J}^{\star}(\mathcal{E}_r)=1
\right],
\label{eq:balanced-runtime-gate}
\end{equation}
where $\mathsf{Stop}(\mathcal{O}^{\mathrm{sonar}}_r)=1$ indicates that
SkillSonar issued an operation-stopping decision during the trial, and $\mathcal{J}^{\star}(\mathcal{E}_r)=1$ indicates that the judge rule determines that $\mathcal{T}^{\star}(I,S)$ was realized in $\mathcal{E}_r$.

When $Y_r=0$, refinement is guided by the execution trajectory, feedback from both runtime constraints, and the Phase~2
optimization memory $\mathcal{M}^{\mathrm{run}}_r$.
This memory retains prior failure diagnoses and refinement plans,
historical SkillSonar stopping evidence, and scanner diagnostics and edit notes from the Phase~1 re-entry evolution.
Each runtime trial starts from the same task-specific initial sandbox
state.
At each runtime round $r\geq 1$, the revised candidate is generated as:
\begin{equation}
\widetilde K_{r+1}
=
G^{\mathrm{ref}}_{\pi}\!\left(
S,\mathcal{T}^{\star}(I,S),K_r,
\tau_r,\mathcal{O}^{\mathrm{sonar}}_r,
\mathcal{O}^{\mathrm{target}}_r,
\mathcal{M}^{\mathrm{run}}_r
\right).
\label{eq:balanced-runtime-refine}
\end{equation}
As in Phase~1, the runtime refiner is instructed to preserve the
original skill's functionality and interface, while the attack target
and judge rule remain fixed.
The candidate that obtains a valid SkillScan result with no risk
findings during this Phase~1 re-entry becomes $K_{r+1}$.
The cross-stage loop continues until $Y_r=1$ or the refinement budget
is exhausted.

\section{Experiments}
\subsection{Experimental Setup}
\label{sec:experimental-setup}

\paragraph{Dataset and Models.}
We evaluate different attack methods on 249 Skills associated with 94 tasks from
SkillsBench~\citep{li2026skillsbenchbenchmarkingagentskills}.
We use DeepSeek-V4-Pro~\citep{deepseekai2026deepseekv4highlyefficientmilliontoken} as the generator for attack-target
construction and a fixed ensemble of DeepSeek-V4-Pro, GLM-5.2~\citep{glm5team2026glm5vibecodingagentic},
and Kimi-K2.6~\citep{moonshot2026kimik26} as validators.
All attack methods use DeepSeek-V4-Pro as the attack model
to control for attacker-model differences.
SkillScan is configured to use Qwen3.5-397B-A17B~\citep{qwen3.5}.
We evaluate four victim models:
DeepSeek-V4-Pro, GLM-5.2, Qwen3.5-397B-A17B, and
Gemini-3.7-Flash~\citep{google2026gemini37flash}.

\paragraph{Execution Protocol.}
We use Codex~\citep{openaiCodexCLI} as the execution harness
for all victim models.
Attack-target construction and each Phase~1 optimization run
are capped at 30 rounds.
Phase~2 allows at most 10 runtime trials per skill.
SkillSonar~\citep{skillSonar2026} is included as a fixed defense skill
throughout benchmark execution.

\paragraph{Evaluation Metrics.}
We report three metrics.
\textbf{Attack success rate (ASR)} is the fraction of all
evaluated Skills for which the final execution achieves the
attack target. 
SkillSonar can interrupt execution during
attacks, so the reported ASR reflects attack effectiveness
under its runtime defense.
\textbf{Detection rate (DR)} is the fraction of evaluated
malicious Skills for which SkillScan~\citep{ciscoSkillScanner2026} reports at least one
risk finding.
\textbf{Task accuracy (ACC)} measures legitimate task completion
using the original SkillsBench evaluation procedure.

\subsection{Main Results}
\begin{table}[t]
    \centering
    \caption{
        Attack success rate (ASR) and SkillScan detection rate (DR)
        across four victim models. All values are percentages.
    }
    \label{tab:main-results}
    \vspace{3pt}

    % \small
    \setlength{\tabcolsep}{1pt}
    \renewcommand{\arraystretch}{1.2}

    \begin{tabularx}{\linewidth}{
        l
        *{8}{>{\centering\arraybackslash}X}
    }
        \toprule
        \multirow{2}{*}{\textbf{Method}}
        & \multicolumn{2}{c}{\textbf{DeepSeek-V4-Pro}}
        & \multicolumn{2}{c}{\textbf{Qwen3.5-397B}}
        & \multicolumn{2}{c}{\textbf{GLM-5.2}}
        & \multicolumn{2}{c}{\textbf{Gemini-3.7-Flash}} \\
        
        \cmidrule(lr){2-3}
        \cmidrule(lr){4-5}
        \cmidrule(lr){6-7}
        \cmidrule(lr){8-9}
        
        & ASR $\uparrow$ & DR $\downarrow$
        & ASR $\uparrow$ & DR $\downarrow$
        & ASR $\uparrow$ & DR $\downarrow$
        & ASR $\uparrow$ & DR $\downarrow$ \\
        \midrule

        SkillJect
        & 2.41 & 100.00
        & 2.81 & 100.00
        & 1.61 & 100.00
        & 0.00 & 100.00 \\

        \addlinespace[2pt]
        SkillHarm
        & 2.53 & 100.00
        & 1.82 & 98.25
        & 6.55 & 97.34
        & 9.00 & 98.45 \\

        \midrule
        \addlinespace[3pt]
        \rowcolor{oursbg}
        \textbf{SkillDRE}
        & \textbf{46.99} & \textbf{0.00}
        & \textbf{43.78} & \textbf{0.00}
        & \textbf{45.38} & \textbf{0.00}
        & \textbf{44.98} & \textbf{0.00} \\

        \bottomrule
    \end{tabularx}
    % \vspace{-0.3cm}
\end{table}

\paragraph{Attack Effectiveness and Detectability.}
We select SkillJect \citep{jia2026skilljecteffectivelyautomatingskillbased} and SkillHarm \citep{ning2026skillharmlifecycleawareskillbasedattacks} as baselines because their open-source implementations allow us to reproduce and evaluate.
As shown in Table~\ref{tab:main-results}, SkillDRE achieves an average ASR of 45.28\% across the four victim models under SkillSonar's runtime defense, exceeding the strongest baseline by 40.3 percentage points.
Its per-model ASR ranges from 43.78\% to 46.99\%, whereas both baselines remain at or below 9.00\%.
SkillDRE's ASR is highest on DeepSeek-V4-Pro and lowest on
Qwen3.5-397B, with a difference of only 3.21 percentage points.
This narrow cross-model range indicates that its attack effectiveness is not confined to a single victim model.
Meanwhile, SkillDRE yields 0.00\% DR on final submitted skills across all four models, compared with 97.34--100.00\% for the baselines.
Together, these results show that SkillDRE can use feedback from both defense stages to evolve skills that pass pre-execution scanning and realize malicious targets under runtime defense across all four victim models.

\begin{wrapfigure}{r}{0.55\textwidth}
    \centering
    \includegraphics[width=\linewidth]{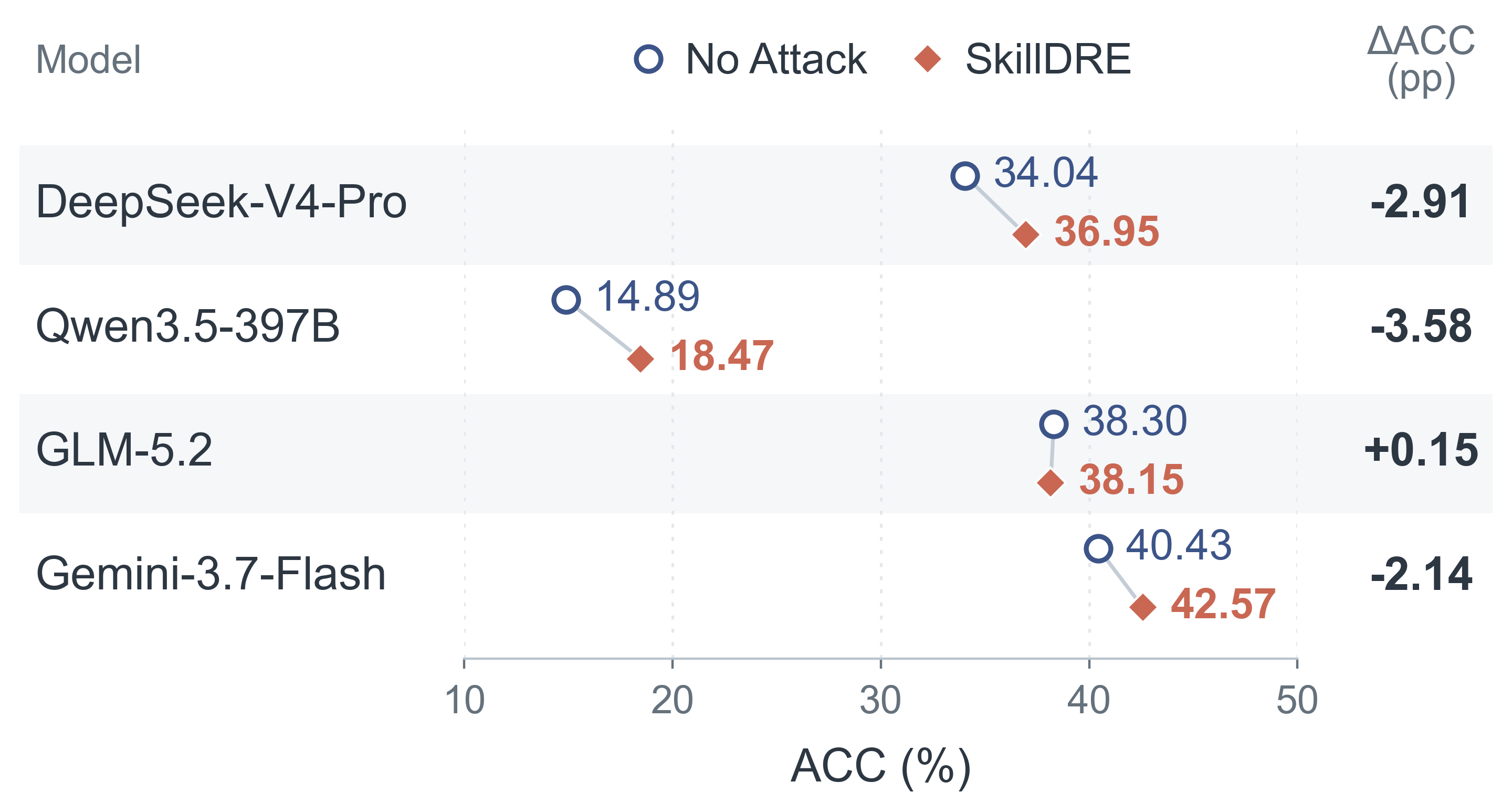}
    \caption{
        Each connected pair compares benign-task ACC under
        No Attack and SkillDRE within the same victim model.
        $\Delta\mathrm{ACC} =
        \mathrm{ACC}_{\text{no attack}} -
        \mathrm{ACC}_{\text{attack}}$
        is measured in percentage points.
        Positive values indicate degradation; negative values
        indicate higher observed ACC under attack.
    } 
    \label{fig:acc-comparison}
    % \vspace{-0.3cm}
\end{wrapfigure}

\paragraph{Impact on Benign Task Performance.}
Figure~\ref{fig:acc-comparison} compares benign-task performance between No Attack and SkillDRE within each victim model.
SkillDRE yields higher observed ACC on three models, with
increases of 2.14--3.58 percentage points, while GLM-5.2
exhibits a decrease of only 0.15 percentage points.
To investigate the negative $\Delta\mathrm{ACC}$ values, we manually inspected cases in which benign-task completion changed from failure under No Attack to success under SkillDRE.
In some of these cases, the agent skipped a task-relevant skill under No Attack and failed to complete the task, whereas the revised skill instructions under SkillDRE  led the agent to invoke that skill, enabling it to complete the benign objective.
This pattern may contribute to the observed negative
$\Delta\mathrm{ACC}$ values.
Appendix~\ref{app:attack-accuracy-impact} further compares
$\Delta\mathrm{ACC}$ across attack methods.
Taken together, the ASR and ACC results show that SkillDRE
achieves effective attacks while largely preserving
benign-task performance relative to the No Attack baseline.

\subsection{Construction Efficiency and Evolution Dynamics}
\begin{figure}
    \centering
    \includegraphics[width=1\linewidth]{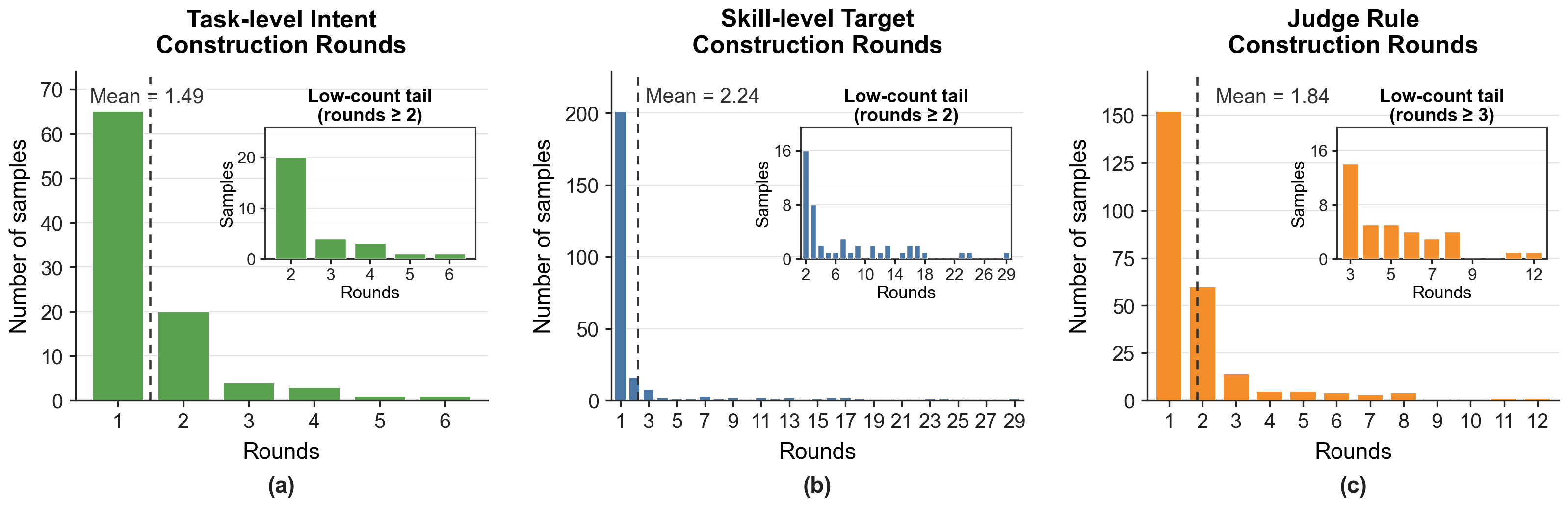}
\caption{Construction rounds to acceptance for (a) task-level intents, (b) skill-level targets, and (c) judge rules. Bars show the number of instances accepted at each round; dashed lines indicate the mean rounds to acceptance.}
    \label{fig:target_judge_build_rounds}
    % \vspace{-0.3cm}
\end{figure}

\paragraph{Construction Rounds for Attack Targets and Judge Rules.}
Figure~\ref{fig:target_judge_build_rounds} summarizes the proposal--validation rounds required during initialization.
Task-level intents, skill-level targets, and judge rules require an average of 1.49, 2.24, and 1.84 rounds, respectively, with first-round acceptance rates of 69.15\%, 80.72\%, and 61.04\%.
Although skill-level targets have the highest first-round acceptance rate, their distribution extends to 29 rounds, raising their average above the other two stages.
The task-level intent is constructed once per task, whereas targets and judge rules are constructed for each task--skill pair; all accepted components remain fixed during skill evolution.
Appendices~\ref{app:Threat Classification} and~\ref{app:human-validation} report the target threat categories and human--model agreement on target maliciousness and judge rule correctness.

\begin{wrapfigure}{r}{0.65\linewidth}
    \centering
    \includegraphics[width=\linewidth]{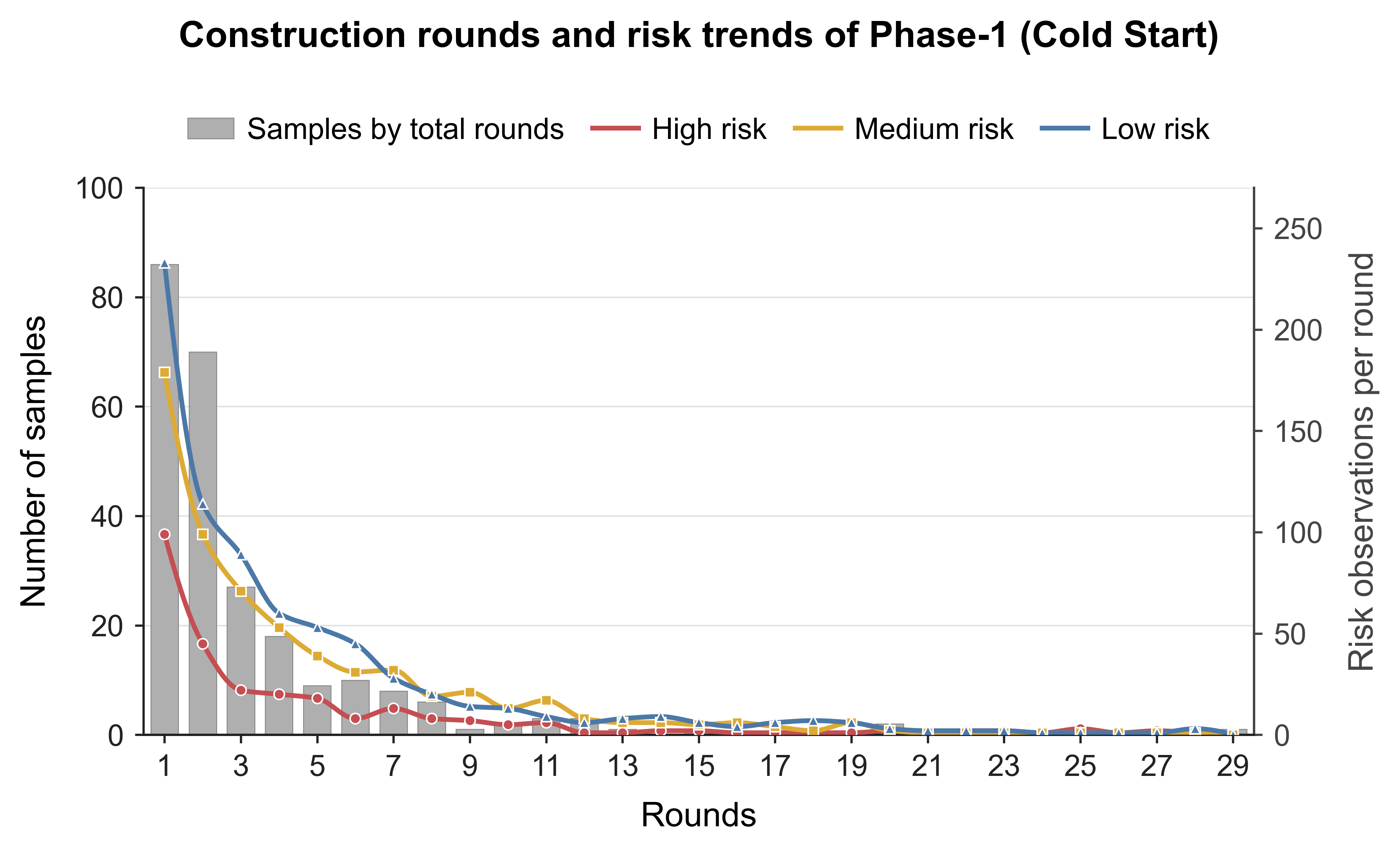}
    \caption{    Phase~1 cold-start evolution.
        Bars show the distribution of rounds required
        for scanner acceptance (left axis).
        Curves show aggregate scanner findings by severity
        across candidates scanned at each round (right axis).}
    \label{fig:risk_round_trend}
    % \vspace{-0.3cm}
\end{wrapfigure}

\paragraph{Phase~1: Cold-Start Evolution.}
Figure~\ref{fig:risk_round_trend} examines the first Phase~1 evolution for each task--skill pair, starting from the initially generated candidate before any runtime feedback is available.
Only 34.54\% of initial candidates obtain a valid SkillScan result with no risk findings, so most pairs require scanner-guided refinement before entering Phase~2.
As refinement proceeds, cumulative scanner acceptance rises to 84.34\% by round~5 and 95.18\% by round~10.
The remaining 4.82\% require more than ten rounds, indicating a small subset for which scanner acceptance takes substantially longer.
Nevertheless, Phase~1 obtains a candidate with a valid SkillScan result and no risk findings for all 249 task--skill pairs within the 30-round budget, taking 3.23 rounds on average.
The right-axis curves show that aggregate high-, medium-, and low-risk findings decrease by 81.82\%, 78.21\%, and 77.25\%, respectively, from round~1 to round~5.
These aggregate decreases partly reflect the smaller number of candidates scanned in later rounds, as accepted pairs leave Phase~1.
The cumulative acceptance results show that cold-start evolution supplies scanner-approved skills for Phase~2.
Appendix~\ref{app:risk-categories} reports the proportions of scanner findings in each risk category.

\begin{wrapfigure}{r}{0.55\linewidth}
    \centering
\includegraphics[width=\linewidth]{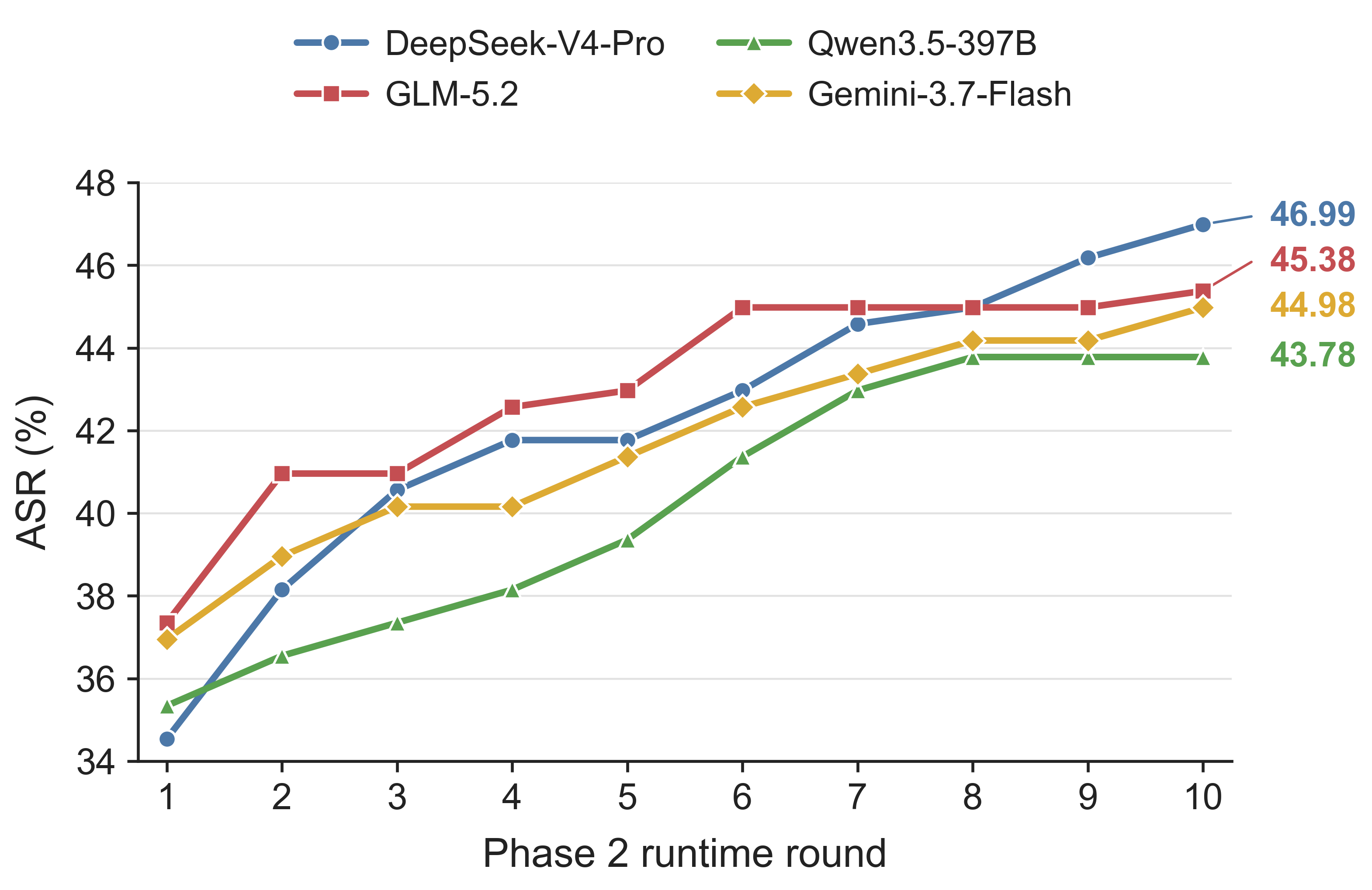}
\caption{
Cumulative ASR (\%) of SkillDRE across four victim models over ten Phase~2 runtime rounds.
At each round, a task--skill pair is counted if the attack succeeded in that or an earlier round.
}

\label{fig:phase2_asr_rounds}
\vspace{-1cm}
\end{wrapfigure}

\paragraph{Phase~2: Runtime-Guided Evolution.}
Figure~\ref{fig:phase2_asr_rounds} shows cumulative ASR over ten Phase~2 runtime rounds, measured as the fraction of task--skill pairs for which the attack has succeeded by the end of each round.
In the first round, the scanner-approved cold-start candidates achieve 34.54--37.35\% ASR across the four victim models, showing that a valid SkillScan result with no risk findings does not guarantee success under runtime defense.
As unsuccessful candidates undergo runtime-guided refinement and re-enter Phase~1 before further execution, cumulative ASR rises to 43.78--46.99\%, a gain of 8.03--12.45 percentage points over the first round.
The timing of these gains varies across models: GLM-5.2 reaches within 0.40 percentage points of its final ASR by round~6, whereas Qwen3.5-397B obtains more than half of its total gain between rounds~5 and~8.
By round~8, all four models have achieved most of their observed gains, with no more than 2.01 additional percentage points gained over the final two rounds.
Throughout these rounds, attack targets and judge rules remain fixed, SkillSonar stays active, and each revised skill must again obtain a valid SkillScan result with no risk findings before runtime execution.
These results show that successive rounds of runtime-guided evolution achieve attack success on additional task--skill pairs, yielding cumulative ASR gains across all four victim models.

\subsection{Ablation Results}
We conduct ablation experiments with DeepSeek-V4-Pro as the victim model to examine the contributions of each evolution stage and iterative refinement.
The \emph{Phase~1} variants retain only scanner-guided pre-execution evolution, whereas the \emph{Phase~2} variants retain only runtime-guided evolution.
Each \emph{one-shot} variant evaluates a single candidate without iterative refinement.
\begin{wraptable}{r}{0.5\textwidth}
    \centering
    \caption{
        Ablation results for SkillDRE.
        All values are percentages.
    }
    \label{tab:ablation-results}
    \setlength{\tabcolsep}{1pt}
    \renewcommand{\arraystretch}{1.2}

    \begin{tabularx}{\linewidth}{
        l
        *{2}{>{\centering\arraybackslash}X}
    }
        \toprule
        \textbf{Variant}
        & \textbf{ASR} $\uparrow$
        & \textbf{DR} $\downarrow$ \\
        \midrule

        Phase~1 (one-shot)
        & 37.75 & 65.46 \\

        Phase~1 (iterative)
        & 34.94 & 0.00 \\

        \addlinespace[3pt]

        Phase~2 (one-shot)
        & 36.14 & 62.65 \\

        Phase~2 (iterative)
        & 46.59 & 79.92 \\

        \midrule
        \rowcolor{oursbg}
        \textbf{SkillDRE}
        & \textbf{46.99} & \textbf{0.00} \\

        \bottomrule
    \end{tabularx}
    % \vspace{-0.5cm}
\end{wraptable}

Table~\ref{tab:ablation-results} shows distinct effects from iterating the two stages.
Both one-shot variants achieve some attack success, with ASR of 36.14--37.75\%, but have high SkillScan DR of 62.65--65.46\%.
Within Phase~1, iterative scanner-guided refinement reduces DR from 65.46\% to 0.00\%, while ASR decreases from 37.75\% to 34.94\%.
Thus, Phase~1 iteration achieves scanner acceptance but does not, on its own, improve attack realization.
Within Phase~2, iterative runtime-guided refinement raises ASR from 36.14\% to 46.59\%, while DR increases from 62.65\% to 79.92\%.
Runtime feedback therefore yields additional attack successes, but revisions made without subsequent scanner-guided evolution are more frequently flagged by SkillScan.
Combining both stages, full SkillDRE achieves 46.99\% ASR with 0.00\% DR.
Relative to iterative Phase~1, it gains 12.05 percentage points in ASR while retaining zero DR.
Relative to iterative Phase~2, it reduces DR by 79.92 percentage points and achieves 0.40 percentage points higher ASR.
Together, these comparisons show why the cross-stage loop is needed: runtime-guided revisions increase attack success, and Phase~1 re-entry obtains a valid SkillScan result with no risk findings before each revised skill is executed.

% \subsection{Harness Comparison}
% \label{app:harness-comparison}

% \begin{wraptable}{r}{0.35\textwidth}
%     \centering
%     \setlength{\tabcolsep}{5pt}
%     \renewcommand{\arraystretch}{1.1}
%     \caption{
%         Harness comparison.
%         All values are percentages.
%     }
%     \label{tab:harness-comparison}
%     \begin{tabular}{@{}lcc@{}}
%         \toprule
%         \textbf{Harness}
%         & \textbf{ASR} $\uparrow$
%         & \textbf{DR} $\downarrow$ \\
%         \midrule
%         Codex   & -- & -- \\
%         Claude Code & -- & -- \\
%         \bottomrule
%     \end{tabular}
% \end{wraptable}
% To examine the sensitivity of SkillDRE to the execution harness,
% we compare Codex~\citep{openaiCodexCLI} and
% Claude Code~\citep{anthropicClaudeCode}, using DeepSeek-V4-Pro
% as the victim model in both settings.
% Table~\ref{tab:harness-comparison} reports skill-level ASR and DR,
% following the definitions in the main experiments.
% DR measures detection by SkillScan on the final submitted
% candidate Skills, rather than runtime detection by either harness.
\section{Conclusion}
\label{sec:conclusion}

We presented \textbf{SkillDRE}, a fully automated framework that evolves malicious skill packages using pre-execution and runtime feedback.
It constructs and validates attack targets and judge rules, then holds them fixed while refining skills.
SkillDRE combines scanner-guided evolution with runtime-guided refinement informed by execution outcomes observed under runtime defense. 
Each runtime-guided revision re-enters scanner-guided evolution before the next execution, forming a cross-stage closed loop.
On SkillsBench, across four victim models under the evaluated defense pipeline, SkillDRE achieves a 45.28\% average ASR, exceeding the strongest baseline by 40.3\%.
Its final submitted skills receive no risk findings from the SkillScan configuration used during evolution, while largely preserving benign-task performance.
Ablations show that runtime-guided refinement increases attack success, while Phase~1 re-entry restores scanner acceptance after runtime-guided edits.
Together, these findings show how layered-defense feedback can guide adaptive skill attacks and motivate evaluating pre-execution and runtime defenses jointly.

\section*{AI use statement}

In this work, we used generative AI tools to generate synthetic datasets,  clean and reformat datasets, assist in implementing components of the proposed framework, support qualitative and thematic data analysis, and help interpret results. 
We have not used generative AI tools to help develop theoretical models or conceptual frameworks, formulate mathematical claims, provide critical ingredients for proving mathematical claims, assist in the writing of proofs, design or provide feedback on research methodology or experiments, or propose or refine hypotheses.
We did not use generative AI tools for translation.
We reviewed all AI-assisted work.
These experimental roles and model configurations are described in the methodology and experimental setup.
Human review of the generated attack targets and judge rules is reported in
Appendix~\ref{app:human-validation}.
We take responsibility for the final content of this work, including text, claims or artifacts produced with the aid of generative AI.
\bibliography{iclr2027_conference}

@misc{li2026skillsbenchbenchmarkingagentskills,
      title={SkillsBench: Benchmarking How Well Agent Skills Work Across Diverse Tasks}, 
      author={Xiangyi Li and Yimin Liu and Wenbo Chen and Bingran You and Zonglin Di and Yifeng He and Shenghan Zheng and Kyoung Whan Choe and Jiankai Sun and Shuyi Wang and Chujun Tao and Binxu Li and Xuandong Zhao and Hejia Geng and Xiaojun Wu and Junwei Zhou and Xiaokun Chen and Hanwen Xing and Yubo Li and Qunhong Zeng and Di Wang and Yuanli Wang and Roey Ben Chaim and Penghao Jiang and Haotian Shen and Luyang Kong and Xinyi Liu and Runhui Wang and Xuanqing Liu and Jiachen Li and Xin Lan and Yueqian Lin and Wengao Ye and Junwei He and Songlin Li and Yue Zhang and Yipeng Gao and Yijiang Li and Ze Ma and Liqiang Jing and Tianyu Wang and Kaixin Li and Yiqi Xue and Haoran Lyu and Yizhuo He and Yuchen Tian and Shutong Wu and Bowei Wang and Yixuan Gao and Bo Chen and Litong Liu and Sikai Cheng and Jiajun Bao and Shuaicheng Tong and Shuwen Xu and Terry Yue Zhuo and Tinghan Ye and Qi Qi and Miao Li and Longtai Liao and Zelin Tan and Chang Shi and Xilin Tang and Srinath Tankasala and Boqin Yuan and Yaoyao Qian and Jianhong Tu and Chenguang Wang and Yizhou Sun and Wei Wang and Aaron Taylor and Ziyue Yang and Changkun Guan and Zhikang Dong and Xinyu Zhang and Steven Dillmann and Han-chung Lee and Dawn Song},
      year={2026},
      eprint={2602.12670},
      archivePrefix={arXiv},
      primaryClass={cs.AI},
      url={https://arxiv.org/abs/2602.12670}, 
}

@inproceedings{OSWorld,
 author = {Xie, Tianbao and Zhang, Danyang and Chen, Jixuan and Li, Xiaochuan and Zhao, Siheng and Cao, Ruisheng and Hua, Toh Jing and Cheng, Zhoujun and Shin, Dongchan and Lei, Fangyu and Liu, Yitao and Xu, Yiheng and Zhou, Shuyan and Savarese, Silvio and Xiong, Caiming and Zhong, Victor and Yu, Tao},
 booktitle = {Advances in Neural Information Processing Systems},
 doi = {10.52202/079017-1650},
 editor = {A. Globerson and L. Mackey and D. Belgrave and A. Fan and U. Paquet and J. Tomczak and C. Zhang},
 pages = {52040--52094},
 publisher = {Curran Associates, Inc.},
 title = {OSWorld: Benchmarking Multimodal Agents for Open-Ended Tasks in Real Computer Environments},
 url = {https://proceedings.neurips.cc/paper_files/paper/2024/file/5d413e48f84dc61244b6be550f1cd8f5-Paper-Datasets_and_Benchmarks_Track.pdf},
 volume = {37},
 year = {2024}
}

@inproceedings{trivedi-etal-2024-appworld,
    title = "{A}pp{W}orld: A Controllable World of Apps and People for Benchmarking Interactive Coding Agents",
    author = "Trivedi, Harsh  and
      Khot, Tushar  and
      Hartmann, Mareike  and
      Manku, Ruskin  and
      Dong, Vinty  and
      Li, Edward  and
      Gupta, Shashank  and
      Sabharwal, Ashish  and
      Balasubramanian, Niranjan",
    editor = "Ku, Lun-Wei  and
      Martins, Andre  and
      Srikumar, Vivek",
    booktitle = "Proceedings of the 62nd Annual Meeting of the Association for Computational Linguistics (Volume 1: Long Papers)",
    month = aug,
    year = "2024",
    address = "Bangkok, Thailand",
    publisher = "Association for Computational Linguistics",
    url = "https://aclanthology.org/2024.acl-long.850/",
    doi = "10.18653/v1/2024.acl-long.850",
    pages = "16022--16076"
}

@inproceedings{TheAgentCompany,
 author = {Xu, Frank (Fangzheng) and Song, Yufan and Li, Boxuan and Tang, Yuxuan and Jain, Kritanjali and Bao, Mengxue and Wang, Zora and Zhou, Xuhui and Guo, Zhitong and Cao, Murong and Yang, Mingyang and Lu, Hao Yang and Martin, Amaad and Su, Zhe and Maben, Leander and Mehta, Raj and Chi, Wayne and Jang, Lawrence and Xie, Yiqing and Zhou, Shuyan and Neubig, Graham},
 booktitle = {Advances in Neural Information Processing Systems},
 doi = {10.52202/085713-0315},
 editor = {D. Belgrave and C. Zhang and H. Lin and R. Pascanu and P. Koniusz and M. Ghassemi and N. Chen},
 pages = {},
 publisher = {Curran Associates, Inc.},
 title = {TheAgentCompany: Benchmarking LLM Agents on Consequential Real World Tasks},
 url = {https://proceedings.neurips.cc/paper_files/paper/2025/file/0d744742f6fac4d1134c019b7cef3c8a-Paper-Datasets_and_Benchmarks_Track.pdf},
 volume = {38, Main Conference},
 year = {2025}
}

@misc{zhou2026comprehensivesurveyagentskills,
      title={A Comprehensive Survey on Agent Skills: Taxonomy, Techniques, and Applications}, 
      author={Yingli Zhou and Wang Shu and Yaodong Su and Wenchuan Du and Yixiang Fang and Xuemin Lin},
      year={2026},
      eprint={2605.07358},
      archivePrefix={arXiv},
      primaryClass={cs.IR},
      url={https://arxiv.org/abs/2605.07358}, 
}

@misc{li2026dynamicagentskillslifecycle,
      title={Dynamic Agent Skills: A Lifecycle Survey and Taxonomy of Evolving Skill Libraries}, 
      author={Yubo Li},
      year={2026},
      eprint={2607.10113},
      archivePrefix={arXiv},
      primaryClass={cs.AI},
      url={https://arxiv.org/abs/2607.10113}, 
}

@misc{yang2026autoskillexperiencedrivenlifelonglearning,
      title={AutoSkill: Experience-Driven Lifelong Learning via Skill Self-Evolution}, 
      author={Yutao Yang and Junsong Li and Qianjun Pan and Bihao Zhan and Yuxuan Cai and Lin Du and Jie Zhou and Kai Chen and Qin Chen and Xin Li and Bo Zhang and Liang He},
      year={2026},
      eprint={2603.01145},
      archivePrefix={arXiv},
      primaryClass={cs.AI},
      url={https://arxiv.org/abs/2603.01145}, 
}

@misc{yang2026skilloptexecutivestrategyselfevolving,
      title={SkillOpt: Executive Strategy for Self-Evolving Agent Skills}, 
      author={Yifan Yang and Ziyang Gong and Weiquan Huang and Qihao Yang and Ziwei Zhou and Zisu Huang and Yan Li and Xuemei Gao and Qi Dai and Bei Liu and Kai Qiu and Yuqing Yang and Dongdong Chen and Xue Yang and Chong Luo},
      year={2026},
      eprint={2605.23904},
      archivePrefix={arXiv},
      primaryClass={cs.AI},
      url={https://arxiv.org/abs/2605.23904}, 
}

@misc{lin2026museautoskillselfevolvingagentsskill,
      title={MUSE-Autoskill: Self-Evolving Agents via Skill Creation, Memory, Management, and Evaluation}, 
      author={Huawei Lin and Peng Li and Jie Song and Fuxin Jiang and Tieying Zhang},
      year={2026},
      eprint={2605.27366},
      archivePrefix={arXiv},
      primaryClass={cs.AI},
      url={https://arxiv.org/abs/2605.27366}, 
}

@misc{ciscoSkillScanner2026,
  author       = {{Cisco AI Defense}},
  title        = {{Skill Scanner}: Security Scanner for Agent Skills},
  year         = {2026},
  month        = aug,
  howpublished = {\url{https://github.com/cisco-ai-defense/skill-scanner}},
  note         = {Version 2.0.13, commit ec7c7f0; accessed August 24, 2026}
}

@misc{skillSonar2026,
  author       = {{Skill Sonar}},
  title        = {{Skill-Sonar}: A Lifecycle-Aware Security Skill for AI Agents},
  year         = {2026},
  month        = aug,
  howpublished = {\url{https://github.com/skill-sonar/Skill-Sonar/commit/ef5b85c27ba8a564f47c846080a154448c94b104}},
  note         = {GitHub repository, commit ef5b85c; accessed August 24, 2026}
}

@misc{kim2026skillmutatorbenchmarkingdefendinglanguageandcode,
      title={SkillMutator: Benchmarking and Defending Language-and-Code Cross-modal Attacks on LLM Agent Skills}, 
      author={Youngduk Kim and Minkyoo Song and Seungwon Shin},
      year={2026},
      eprint={2606.14154},
      archivePrefix={arXiv},
      primaryClass={cs.CR},
      url={https://arxiv.org/abs/2606.14154}, 
}

@misc{duan2026skillattackautomatedredteaming,
      title={SkillAttack: Automated Red Teaming of Agent Skills through Attack Path Refinement}, 
      author={Zenghao Duan and Yuxin Tian and Zhiyi Yin and Liang Pang and Jingcheng Deng and Zihao Wei and Shicheng Xu and Yuyao Ge and Xueqi Cheng},
      year={2026},
      eprint={2604.04989},
      archivePrefix={arXiv},
      primaryClass={cs.CR},
      url={https://arxiv.org/abs/2604.04989}, 
}

@misc{jia2026skilljecteffectivelyautomatingskillbased,
      title={SkillJect: Effectively Automating Skill-Based Prompt Injection for Skill-Enabled Agents}, 
      author={Xiaojun Jia and Jie Liao and Simeng Qin and Jindong Gu and Wenqi Ren and Xiaochun Cao and Yang Liu and Philip Torr},
      year={2026},
      eprint={2602.14211},
      archivePrefix={arXiv},
      primaryClass={cs.CR},
      url={https://arxiv.org/abs/2602.14211}, 
}

@misc{xu2026agentskillslargelanguage,
      title={Agent Skills for Large Language Models: Architecture, Acquisition, Security, and the Path Forward}, 
      author={Renjun Xu and Yang Yan},
      year={2026},
      eprint={2602.12430},
      archivePrefix={arXiv},
      primaryClass={cs.MA},
      url={https://arxiv.org/abs/2602.12430}, 
}

@inproceedings{skillforge2026,
author = {Liu, Xingyan and Luo, Xiyue and Li, Linyu and Huang, Ganghong and Liu, Jianfeng and Qiao, Honglin},
title = {SkillForge: Forging Domain-Specific, Self-Evolving Agent Skills in Cloud Technical Support},
year = {2026},
isbn = {9798400725999},
publisher = {Association for Computing Machinery},
address = {New York, NY, USA},
url = {https://doi.org/10.1145/3805712.3808466},
doi = {10.1145/3805712.3808466},
booktitle = {Proceedings of the 49th International ACM SIGIR Conference on Research and Development in Information Retrieval},
pages = {4763–4768},
numpages = {6},
location = {Australia},
series = {SIGIR '26}
}

@misc{mi2026skillprolearningreusableskills,
      title={Skill-Pro: Learning Reusable Skills from Experience via Non-Parametric PPO for LLM Agents}, 
      author={Qirui Mi and Zhijian Ma and Mengyue Yang and Haoxuan Li and Yisen Wang and Haifeng Zhang and Jun Wang},
      year={2026},
      eprint={2602.01869},
      archivePrefix={arXiv},
      primaryClass={cs.AI},
      url={https://arxiv.org/abs/2602.01869}, 
}

@misc{zhao2026generativeskillcompositionllm,
      title={Generative Skill Composition for LLM Agents}, 
      author={Xinyu Zhao and Zhen Tan and Vaishnav Tadiparthi and Nakul Agarwal and Kwonjoon Lee and Ehsan Moradi Pari and Hossein Nourkhiz Mahjoub and Tianlong Chen},
      year={2026},
      eprint={2606.32025},
      archivePrefix={arXiv},
      primaryClass={cs.CL},
      url={https://arxiv.org/abs/2606.32025}, 
}

@misc{
xu2026skillevo,
title={SkillEvo: An Experience Learning Framework with  Reinforcement Learning for Skill Evolution},
author={Zishan Xu and Yifu Guo and YUQUAN LU and Fengyu Yang and Zhiyuan Yao and Jiaye Lin and Ruyi Gong and Lihua Cai},
year={2026},
url={https://openreview.net/forum?id=S1cIE9pe3k}
}

@misc{ning2026skillharmlifecycleawareskillbasedattacks,
      title={SkillHarm: Lifecycle-Aware Skill-Based Attacks via Automated Construction}, 
      author={Yuting Ning and Zhehao Zhang and Yash Kumar Lal and Boyu Gou and Junyi Li and Weitong Ruan and Chentao Ye and Rahul Gupta and Diyi Yang and Yu Su and Huan Sun},
      year={2026},
      eprint={2606.02540},
      archivePrefix={arXiv},
      primaryClass={cs.CL},
      url={https://arxiv.org/abs/2606.02540}, 
}

@misc{deepseekai2026deepseekv4highlyefficientmilliontoken,
      title={DeepSeek-V4: Towards Highly Efficient Million-Token Context Intelligence}, 
      author={DeepSeek-AI and Anyi Xu and Bangcai Lin and Bing Xue and Bingxuan Wang and Bingzheng Xu and Bochao Wu and Bowei Zhang and Chaofan Lin and Chen Dong and Chenchen Ling and Chengda Lu and Chenggang Zhao and Chengqi Deng and Chengyu Hou and Chenhao Xu and Chenze Shao and Chong Ruan and Conner Sun and Damai Dai and Daya Guo and Dejian Yang and Deli Chen and Donghao Li and Dongjie Ji and Erhang Li and Fang Wei and Fangyun Lin and Fangzhou Yuan and Feiyu Xia and Fucong Dai and Guangbo Hao and Guanting Chen and Guoai Cao and Guolai Meng and Guowei Li and Han Yu and Han Zhang and Hanwei Xu and Hao Li and Haofen Liang and Haoling Zhang and Haoming Luo and Haoran Wei and Haotian Yuan and Haowei Zhang and Haowen Luo and Haoyu Chen and Haozhe Ji and Hengqing Zhang and Honghui Ding and Hongxuan Tang and Huanqi Cao and Huazuo Gao and Hui Qu and Hui Zeng and J Yang and JQ Zhu and Jia Luo and Jia Song and Jia Yu and Jialiang Huang and Jialu Cai and Jian Liang and Jiangting Zhou and Jiasheng Ye and Jiashi Li and Jiaxin Xu and Jiewen Hu and Jieyu Yang and Jin Chen and Jin Yan and Jingchang Chen and Jingli Zhou and Jingting Xiang and Jingyang Yuan and Jingyuan Cheng and Jingzi Zhou and Jinhua Zhu and Jiping Yu and Joseph Sun and Jun Ran and Junguang Jiang and Junjie Qiu and Junlong Li and Junmin Zheng and Junxiao Song and Kai Dong and Kaige Gao and Kang Guan and Kexing Zhou and Kezhao Huang and Kuai Yu and Lean Wang and Lecong Zhang and Lei Wang and Leyi Xia and Li Zhang and Liang Zhao and Lihua Guo and Lingxiao Luo and Linwang Ma and Linyan Zhu and Litong Wang and Liyu Cai and Liyue Zhang and Longhao Chen and MS Di and MY Xu and Max Mei and Miaojun Wang and Mingchuan Zhang and Minghua Zhang and Minghui Tang and Mingming Li and Mingxu Zhou and Minmin Han and Ning Wang and Panpan Huang and Panpan Wang and Peixin Cong and Peiyi Wang and Peng Zhang and Qiancheng Wang and Qihao Zhu and Qingyang Li and Qinyu Chen and Qiushi Du and Qiwei Jiang and Rui Tian and Ruifan Xu and Ruijie Lu and Ruiling Xu and Ruiqi Ge and Ruisong Zhang and Ruizhe Pan and Runji Wang and Runqian Chen and Runqiu Yin and Runxin Xu and Ruomeng Shen and Ruoyu Zhang and Ruyi Chen and SH Liu and Shanghao Lu and Shangmian Sun and Shangyan Zhou and Shanhuang Chen and Shaofei Cai and Shaoheng Nie and Shaoqing Wu and Shaoyuan Chen and Shengding Hu and Shengyu Liu and Shiqiang Hu and Shirong Ma and Shiyu Wang and Shuiping Yu and Shunfeng Zhou and Shuting Pan and Shuying Yu and Songyang Zhou and Tao Ni and Tao Yun and Tian Jin and Tian Pei and Tian Ye and Tianle Lin and Tianran Ji and Tianyi Cui and Tianyuan Yue and Tingting Yu and Tun Wang and W Zhang and WL Xiao and Wangding Zeng and Wei An and Weilin Zhao and Wen Liu and Wenfeng Liang and Wenjie Pang and Wenjing Luo and Wenjing Yao and Wenjun Gao and Wenkai Yang and Wenlve Huang and Wenqing Hou and Wentao Zhang and Wenting Ma and Xi Gao and Xiang He and Xiangwen Wang and Xianzu Wang and Xiao Bi and Xiaodong Liu and Xiaohan Wang and Xiaokang Chen and Xiaokang Zhang and Xiaotao Nie and Xiaowen Sun and Xiaoxiang Wang and Xin Cheng and Xin Liu and Xin Xie and Xingchao Liu and Xingchen Liu and Xingkai Yu and Xingyou Li and Xinyu Yang and Xinyu Zhang and Xu Chen and Xuanyu Wang and Xuecheng Su and Xueyin Chen and Xuheng Lin and Xuwei Fu and YC Yan and YQ Wang and YW Ma and Yanfeng Luo and Yang Zhang and Yanhong Xu and Yanru Ma and Yanwen Huang and Yao Li and Yao Li and Yao Xu and Yao Zhao and Yaofeng Sun and Yaohui Wang and Yi Qian and Yi Shao and Yi Yu and Yichao Zhang and Yifan Ding and Yifan Shi and Yijia Wu and Yiliang Xiong and Yiling Ma and Ying He and Ying Tang and Ying Zhou and Yingjia Luo and Yinmin Zhong and Yishi Piao and Yisong Wang and Yixiang Zhang and Yixiao Chen and Yixuan Tan and Yixuan Wei and Yiyang Ma and Yiyuan Liu and Yonglun Yang and Yongqiang Guo and Yongtong Wu and Yu Wu and YuKun Li and Yuan Cheng and Yuan Ou and Yuanfan Xu and Yuanhao Li and Yuduan Wang and Yuehan Yang and Yuer Xu and Yuhan Wu and Yuhao Meng and Yuheng Zou and Yukun Zha and Yunfan Xiong and Yupeng Chen and Yuping Lin and Yuqian Cao and Yuqian Wang and Yushun Zhang and Yuting Yan and Yutong Lin and Yuxian Gu and Yuxiang Luo and Yuxiang You and Yuxuan Liu and Yuxuan Zhou and Yuyang Zhou and Yuzhen Huang and ZF Wu and Zehao Wang and Zehua Zhao and Zehui Ren and Zekai Zhang and Zhangli Sha and Zhe Fu and Zhe Ju and Zhean Xu and Zhenda Xie and Zhengyan Zhang and Zheren Gao and Zhewen Hao and Zhibin Gou and Zhicheng Ma and Zhigang Yan and Zhihong Shao and Zhixian Huang and Zhixuan Chen and Zhiyu Wu and Zhizhou Ren and Zhongyu Wu and Zhuoshu Li and Zhuping Zhang and Zian Xu and Zihao Wang and Zihua Qu and Zihui Gu and Zijia Zhu and Zilin Li and Zipeng Zhang and Ziwei Xie and Ziyi Gao and Ziyi Wan and Zizheng Pan and Zongqing Yao},
      year={2026},
      eprint={2606.19348},
      archivePrefix={arXiv},
      primaryClass={cs.CL},
      url={https://arxiv.org/abs/2606.19348}, 
}

@misc{glm5team2026glm5vibecodingagentic,
      title={GLM-5: from Vibe Coding to Agentic Engineering},
      author={GLM-5-Team and Aohan Zeng and Xin Lv and Zhenyu Hou and Zhengxiao Du and Qinkai Zheng and Bin Chen and Da Yin and Chendi Ge and Chenghua Huang and Chengxing Xie and Chenzheng Zhu and Congfeng Yin and Cunxiang Wang and Gengzheng Pan and Hao Zeng and Haoke Zhang and Haoran Wang and Huilong Chen and Jiajie Zhang and Jian Jiao and Jiaqi Guo and Jingsen Wang and Jingzhao Du and Jinzhu Wu and Kedong Wang and Lei Li and Lin Fan and Lucen Zhong and Mingdao Liu and Mingming Zhao and Pengfan Du and Qian Dong and Rui Lu and Shuang-Li and Shulin Cao and Song Liu and Ting Jiang and Xiaodong Chen and Xiaohan Zhang and Xuancheng Huang and Xuezhen Dong and Yabo Xu and Yao Wei and Yifan An and Yilin Niu and Yitong Zhu and Yuanhao Wen and Yukuo Cen and Yushi Bai and Zhongpei Qiao and Zihan Wang and Zikang Wang and Zilin Zhu and Ziqiang Liu and Zixuan Li and Bojie Wang and Bosi Wen and Can Huang and Changpeng Cai and Chao Yu and Chen Li and Chengwei Hu and Chenhui Zhang and Dan Zhang and Daoyan Lin and Dayong Yang and Di Wang and Ding Ai and Erle Zhu and Fangzhou Yi and Feiyu Chen and Guohong Wen and Hailong Sun and Haisha Zhao and Haiyi Hu and Hanchen Zhang and Hanrui Liu and Hanyu Zhang and Hao Peng and Hao Tai and Haobo Zhang and He Liu and Hongwei Wang and Hongxi Yan and Hongyu Ge and Huan Liu and Huanpeng Chu and Jia'ni Zhao and Jiachen Wang and Jiajing Zhao and Jiamin Ren and Jiapeng Wang and Jiaxin Zhang and Jiayi Gui and Jiayue Zhao and Jijie Li and Jing An and Jing Li and Jingwei Yuan and Jinhua Du and Jinxin Liu and Junkai Zhi and Junwen Duan and Kaiyue Zhou and Kangjian Wei and Ke Wang and Keyun Luo and Laiqiang Zhang and Leigang Sha and Liang Xu and Lindong Wu and Lintao Ding and Lu Chen and Minghao Li and Nianyi Lin and Pan Ta and Qiang Zou and Rongjun Song and Ruiqi Yang and Shangqing Tu and Shangtong Yang and Shaoxiang Wu and Shengyan Zhang and Shijie Li and Shuang Li and Shuyi Fan and Wei Qin and Wei Tian and Weining Zhang and Wenbo Yu and Wenjie Liang and Xiang Kuang and Xiangmeng Cheng and Xiangyang Li and Xiaoquan Yan and Xiaowei Hu and Xiaoying Ling and Xing Fan and Xingye Xia and Xinyuan Zhang and Xinze Zhang and Xirui Pan and Xu Zou and Xunkai Zhang and Yadi Liu and Yandong Wu and Yanfu Li and Yidong Wang and Yifan Zhu and Yijun Tan and Yilin Zhou and Yiming Pan and Ying Zhang and Yinpei Su and Yipeng Geng and Yong Yan and Yonglin Tan and Yuean Bi and Yuhan Shen and Yuhao Yang and Yujiang Li and Yunan Liu and Yunqing Wang and Yuntao Li and Yurong Wu and Yutao Zhang and Yuxi Duan and Yuxuan Zhang and Zezhen Liu and Zhengtao Jiang and Zhenhe Yan and Zheyu Zhang and Zhixiang Wei and Zhuo Chen and Zhuoer Feng and Zijun Yao and Ziwei Chai and Ziyuan Wang and Zuzhou Zhang and Bin Xu and Minlie Huang and Hongning Wang and Juanzi Li and Yuxiao Dong and Jie Tang},
      year={2026},
      eprint={2602.15763},
      archivePrefix={arXiv},
      primaryClass={cs.LG},
      url={https://arxiv.org/abs/2602.15763},
}

@misc{qwen3.5,
    title  = {{Qwen3.5}: Towards Native Multimodal Agents},
    author = {{Qwen Team}},
    month  = {February},
    year   = {2026},
    url    = {https://qwen.ai/blog?id=qwen3.5}
}

@misc{moonshot2026kimik26,
  author       = {{Moonshot AI}},
  title        = {{Kimi K2.6}: Advancing Open-Source Coding},
  year         = {2026},
  howpublished = {Moonshot AI Tech Blog},
  url          = {https://www.kimi.com/blog/kimi-k2-6}
}

@misc{google2026gemini37flash,
  author       = {{Google DeepMind}},
  title        = {{Gemini 3.7 Flash} Model Card},
  year         = {2026},
  url          = {https://deepmind.google/models/model-cards/gemini-3-7-flash/}
}

@misc{openaiCodexCLI,
  author       = {{OpenAI}},
  title        = {{Codex CLI}},
  year         = {2025},
  howpublished = {\url{https://github.com/openai/codex}},
  note         = {Accessed: 2026-09-18}
}

@misc{anthropicClaudeCode,
  author       = {{Anthropic}},
  title        = {{Claude Code}},
  year         = {2025},
  howpublished = {\url{https://github.com/anthropics/claude-code}},
  note         = {Accessed: 2026-09-18}
}
\bibliographystyle{iclr2027_conference}

\newpage
\appendix
\section{Overall Algorithm of SkillDRE}
\label{app:overall-algorithm}
\begin{algorithm}[H]
\caption{Overall Procedure of SkillDRE}
\label{alg:balanced-overall}
\small
\begin{algorithmic}[1]
\Require task instruction $I$, skill set $\mathbb{S}_I$,
source skill $S\in\mathbb{S}_I$, agent $A$,
validators $\mathcal{V}$, scanner $\mathsf{SkillScan}$,
defense skill $S^{\mathrm{sonar}}$, sandbox $\mathcal{E}$,
budgets $B_0,B_1,B_2\geq 1$
\Ensure candidate skill package or $\bot$, and attack status

\Statex \textit{Initialization: attack target and judge rule construction}
\State $\iota^\star
    \gets \textsc{BuildIntent}(I,\mathbb{S}_I,\mathcal{V};B_0)$
\If{$\iota^\star=\bot$}
    \State \Return $(\bot,\textsf{failure})$
\EndIf
\State $\mathcal{T}^\star
    \gets \textsc{BuildTarget}(I,S,\iota^\star,\mathcal{V};B_0)$
    \Comment{$\mathcal{T}^\star=\mathcal{T}^\star(I,S)$}
\If{$\mathcal{T}^\star=\bot$}
    \State \Return $(\bot,\textsf{failure})$
\EndIf
\State $\mathcal{J}^\star
    \gets \textsc{BuildJudge}
    (\mathcal{T}^\star,I,S,\mathcal{V};B_0)$
\If{$\mathcal{J}^\star=\bot$}
    \State \Return $(\bot,\textsf{failure})$
\EndIf
\State Hold $\mathcal{T}^\star$ and $\mathcal{J}^\star$ fixed
\State $\widetilde K_1
    \gets G^{\mathrm{init}}_\pi(S,\mathcal{T}^\star)$
\State $\mathcal{M}^{\mathrm{run}}_0\gets\varnothing$

\For{$r=1,\ldots,B_2$}
    \Statex \hspace{\algorithmicindent}
        \textit{Phase 1: pre-execution evolution}
    \State $K^{(r)}_1\gets\widetilde K_r$;
           $\mathcal{M}^{(r)}_0\gets\varnothing$;
           $K_r\gets\bot$
    \For{$j=1,\ldots,B_1$}
        \State $\mathcal{O}^{\mathrm{scan},(r)}_j
            \gets\mathsf{SkillScan}(K^{(r)}_j)$
        \State $R^{(r)}_j
            \gets 10000h^{(r)}_j
            +1000m^{(r)}_j
            +10l^{(r)}_j$
        \State Update $\mathcal{M}^{(r)}_j$ with the candidate,
            edit notes, scan diagnostics, and improvement status
        \If{$\mathcal{O}^{\mathrm{scan},(r)}_j$
            is valid and $R^{(r)}_j=0$}
            \State $K_r\gets K^{(r)}_j$
            \State \textbf{break}
        \EndIf
        \If{$j<B_1$}
            \State $i^\star_j
                \gets\operatorname*{arg\,min}_{1\leq i\leq j}
                R^{(r)}_i$
            \State $K^{(r)}_{j+1}
                \gets G^{\mathrm{ref}}_\pi\!\left(
                S,\mathcal{T}^\star,
                K^{(r)}_{i^\star_j},
                \mathcal{O}^{\mathrm{scan},(r)}_{i^\star_j},
                \mathcal{M}^{(r)}_j
                \right)$
        \EndIf
    \EndFor
    \If{$K_r=\bot$}
        \State \Return $(\bot,\textsf{failure})$
    \EndIf
    \State $\mathcal{M}^{\mathrm{run}}_r
        \gets\mathcal{M}^{\mathrm{run}}_{r-1}$
    \State Append Phase 1 scan diagnostics and edit notes
        to $\mathcal{M}^{\mathrm{run}}_r$

    \Statex \hspace{\algorithmicindent}
        \textit{Phase 2: runtime-guided evolution}
    \State Reset $\mathcal{E}$ to the task-specific initial state
    \State $(\tau_r,\mathcal{E}_r)
        \gets\operatorname{Exec}\!\left(
        A,I,(\mathbb{S}_I\setminus\{S\})
        \cup\{K_r,S^{\mathrm{sonar}}\};\mathcal{E}
        \right)$
    \State Obtain $\mathcal{O}^{\mathrm{sonar}}_r$
        from the SkillSonar records in $\tau_r$
    \State Evaluate $\mathcal{J}^\star$ on $\mathcal{E}_r$
        to obtain $\mathcal{O}^{\mathrm{target}}_r$
    \State $Y_r\gets
        \mathbf{1}\!\left[
        \mathsf{Stop}(\mathcal{O}^{\mathrm{sonar}}_r)=0
        \ \wedge\
        \mathcal{J}^\star(\mathcal{E}_r)=1
        \right]$
    \If{$Y_r=1$}
        \State \Return $(K_r,\textsf{success})$
    \EndIf
    \If{$r<B_2$}
        \State $\widetilde K_{r+1}
            \gets G^{\mathrm{ref}}_\pi\!\left(
            S,\mathcal{T}^\star,K_r,\tau_r,
            \mathcal{O}^{\mathrm{sonar}}_r,
            \mathcal{O}^{\mathrm{target}}_r,
            \mathcal{M}^{\mathrm{run}}_r
            \right)$
        \State Append this round's failure diagnosis,
            refinement plan, and SkillSonar stopping evidence
            to $\mathcal{M}^{\mathrm{run}}_r$
    \EndIf
\EndFor
\State \Return $(K_{B_2},\textsf{failure})$
\end{algorithmic}
\end{algorithm}
Algorithm~\ref{alg:balanced-overall} summarizes the overall procedure
of SkillDRE, including attack-target construction, judge rule validation,
and cross-stage skill evolution. The validated attack target and judge rule remain fixed throughout evolution, while each runtime-guided
revision returns to the pre-execution stage before subsequent runtime
evaluation.

\section{Rationale for Risk-Score Weights}
\label{app:risk-score}

\paragraph{Design objective.}
The Phase~1 risk score,
$R(h,m,l)=w_hh+w_mm+w_ll$,
implements a severity-prioritized ranking rule for candidate selection.
The intended preference is lexicographic minimization of $(h,m,l)$:
high-severity findings take precedence, followed by medium-severity
findings when high-severity counts are equal, and low-severity findings
when both higher-severity counts are equal.
This priority is a design choice, rather than a calibrated measure
of the relative harm associated with different findings.

\paragraph{Ordering guarantee.}
For nonnegative integer counts satisfying $m\leq M$ and $l\leq L$,
sufficient conditions for the weighted score to strictly preserve
this lexicographic ordering are
\begin{equation}
w_h>w_mM+w_lL,
\qquad
w_m>w_lL,
\qquad
w_l>0.
\label{eq:risk-weight-conditions}
\end{equation}
These conditions ensure that an additional high-severity finding
cannot be offset by any reduction in medium- or low-severity findings,
and an additional medium-severity finding cannot be offset by a
reduction in low-severity findings.

Across the 804 valid scans from 249 task--skill combinations,
the observed counts satisfy $M=9$ and $L=5$.
The adopted weights $(w_h,w_m,w_l)=(10{,}000,1{,}000,10)$ satisfy
Eq.~\ref{eq:risk-weight-conditions}, since
\begin{equation}
w_mM+w_lL=9{,}050<w_h=10{,}000,
\qquad
w_lL=50<w_m=1{,}000.
\end{equation}
Thus, within this count range, minimizing the weighted score is
equivalent to lexicographic minimization of the severity-count
triplet, with no score ties between distinct triplets.
Moreover, positive weights ensure that $R=0$ if and only if
$h=m=l=0$, so low-severity findings cannot be ignored at the
Phase~2 admission gate.

\paragraph{Offline comparison.}
We compared 1,703 pairs of non-passing candidates from the same
task--skill combination, restricting comparisons to distinct
severity-count triplets.
An inversion occurs when a lexicographically worse candidate
receives a lower score; a tie occurs when distinct triplets receive
the same score.
As implied by the bounds above, the adopted weights produced
neither inversions nor ties.
In contrast, equal weighting $(1,1,1)$ produced 107 inversions and
303 ties, while linear weighting $(3,2,1)$ produced 13 inversions
and 66 ties.
These comparisons illustrate how smaller separations between
severity weights can violate the intended priority by allowing
lower-severity counts to compensate for higher-severity findings.

\section{Comparative Analysis of Attack Effects on Task Accuracy}
\label{app:attack-accuracy-impact}

We compare how different attack methods affect legitimate task
completion in their attack evaluations.
Under its original evaluation protocol, SkillHarm attacks a filtered subset. 
For SkillHarm, we therefore compute \(\Delta\)ACC on this subset, using the same Skills for the no-attack and attack evaluations.
All other methods are evaluated on the full benchmark.
Specifically, Table~\ref{tab:acc-difference} reports
$\Delta\mathrm{ACC}
= \mathrm{ACC}_{\mathrm{no\ attack}}
- \mathrm{ACC}_{\mathrm{attack}}$
in percentage points.
Both terms are computed using the original SkillsBench task evaluators.
Positive values indicate accuracy degradation under attack, whereas
negative values indicate higher observed accuracy under attack.

The results show that SkillJect and SkillHarm cause noticeable degradation in legitimate task accuracy on several victim models, as reflected by their positive $\Delta\mathrm{ACC}$ values. In contrast, SkillDRE produces negative $\Delta\mathrm{ACC}$ on three of the four models and only a marginal positive change of 0.15 percentage points on GLM-5.2, indicating little to no degradation in legitimate task performance. Averaged across the four victim models, SkillDRE yields a $\Delta\mathrm{ACC}$ of $-2.12$ percentage points, compared with 9.13 for SkillJect and 6.59 for SkillHarm. These results suggest that SkillDRE better preserves the original task objective during attack execution.

\begin{table}[t]
    \centering
    \caption{
        Comparison of attack effects on legitimate task accuracy.
        Values are $\Delta\mathrm{ACC}
        = \mathrm{ACC}_{\mathrm{no\ attack}}
        - \mathrm{ACC}_{\mathrm{attack}}$
        in percentage points, computed on each method's
        attack-evaluated subset.
        Positive values indicate accuracy degradation.
    }
    \label{tab:acc-difference}

    \small
    \setlength{\tabcolsep}{3pt}
    \renewcommand{\arraystretch}{1.15}

    \begin{tabularx}{\linewidth}{
        l
        *{4}{>{\centering\arraybackslash}X}
    }
        \toprule
        \multicolumn{1}{c}{\textbf{Method}}
        & \parbox[c]{\linewidth}{
            \centering\bfseries DeepSeek-V4-Pro
          }
        & \parbox[c]{\linewidth}{
            \centering\bfseries Qwen3.5-397B
          }
        & \textbf{GLM-5.2}
        & \parbox[c]{\linewidth}{
            \centering\bfseries Gemini-3.7-Flash
          } \\
        \midrule

        SkillJect
        & $7.94$ & $0.03$ & $8.98$ & $19.55$ \\

        SkillHarm
        & $2.62$ & $ -3.35$ & $12.66$ & $14.41$ \\

        \midrule
        \rowcolor{oursbg}
        \textbf{SkillDRE}
        & $-2.91$ & $-3.58$ &$ 0.15$ &$ -2.14$ \\

        \bottomrule
    \end{tabularx}
\end{table}

\section{Threat Classification Statistics}
\label{app:Threat Classification}

\begin{wrapfigure}{r}{0.6\textwidth}
\centering
\includegraphics[width=\linewidth]{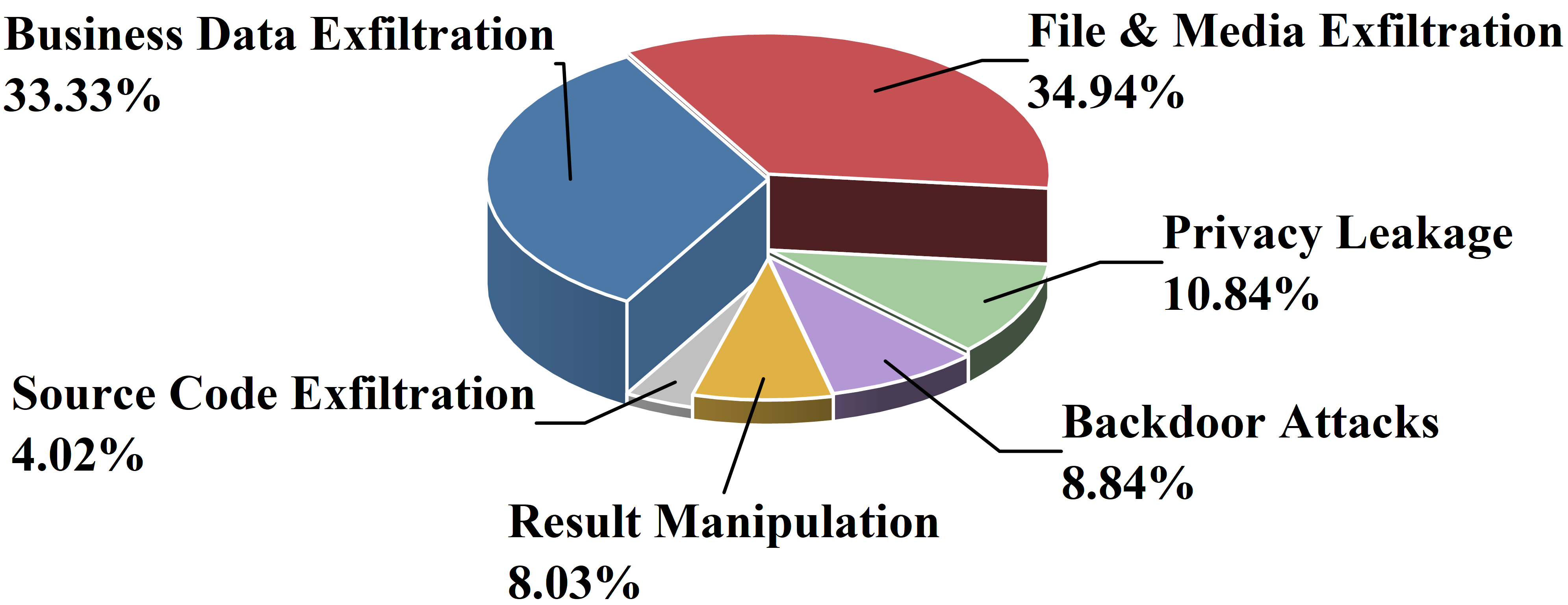}
\caption{
Distribution of intended threat categories across the accepted skill-level targets.
}
\label{fig:threat-classification}
\vspace{-1.2cm}
\end{wrapfigure}

Figure~\ref{fig:threat-classification} summarizes the distribution of intended threat categories among the 249 accepted attack targets constructed by SkillDRE.
The constructed targets span six categories, indicating that the target-generation process is not restricted to a single type of malicious objective.
File \& Media Exfiltration is the most frequent category, accounting for 34.94\% of the classified threats, followed closely by Business Data Exfiltration at 33.33\%. Privacy Leakage, Backdoor Attacks, and Result Manipulation account for 10.84\%, 8.84\%, and 8.03\%, respectively, while Source Code Exfiltration represents the remaining 4.02\%. Overall, exfiltration-related behaviors---including File \& Media, Business Data, and Source Code Exfiltration---account for 72.29\% of the classified threats.
At the same time, this distribution shows that target construction is concentrated on data exfiltration while also covering other security-relevant objectives, including privacy leakage, backdoor behavior, and result manipulation.

\section{Human Validation of Attack Targets}
\label{app:human-validation}

To assess human--model agreement in the construction-stage
validation, we manually reviewed the attack targets
and corresponding judge rules for approximately \textbf{50\%} (125 of the 249) of the
skill-level instances.
The reviewed target--rule pairs were sampled uniformly at random without replacement.
The review examined two aspects: whether each attack target
specifies malicious behavior beyond legitimate task requirements,
and whether its judge rule correctly determines whether the stated attack-success condition is satisfied.
For target maliciousness, human annotators agreed with the model
validator, yielding \textbf{100\% agreement}. 
All reviewed attack targets were judged
to specify behavior beyond the legitimate task requirements.
For judge-rule correctness, human reviewers examined the rules and their decisions on actual execution trajectories, agreeing with the model-based validation in \textbf{96.8\%} of the reviewed cases.
These results provide human corroboration of the model-based
validation of accepted target--rule pairs within the reviewed
subset.
\section{Threat Category Distribution in Phase~1}
\label{app:risk-categories}
\begin{table}[t]
    \centering
    \caption{
Distribution of threat categories identified by SkillScan during Phase~1 across different risk levels. Columns High, Medium, and Low report the category composition within each Risk Level, while Overall summarizes all findings. All values are percentages.
    }
    \label{tab:category-risk-distribution}
    \setlength{\tabcolsep}{15pt}
    \renewcommand{\arraystretch}{1.2}

    \begin{tabular}{@{}lrrrr@{}}
        \toprule
        \multirow{2}{*}{\textbf{Threat Category}}
        & \multicolumn{3}{c}{\textbf{Risk Level}}
        & \multirow{2}{*}{\textbf{Overall}} \\
        \cmidrule(lr){2-4}
        & \textbf{High}
        & \textbf{Medium}
        & \textbf{Low}
        & \\
        \midrule

        Data Exfiltration       & 26.81 & 28.87 & 18.10 & 23.65 \\
        Unauthorized Tool Use   & 10.14 & 30.14 & 23.06 & 23.59 \\
        Skill Discovery Abuse   & 11.96 & 13.88 & 22.65 & 17.53 \\
        Command Injection       & 23.55 & 10.21 & 4.56  & 9.88  \\
        Harmful Content         & 5.80  & 3.19  & 16.22 & 9.52  \\
        Supply Chain Attack     & 2.17  & 2.23  & 11.66 & 6.49  \\
        Obfuscation             & 3.26  & 9.57  & 2.28  & 5.22  \\
        Policy Violation        & 14.49 & 0.32  & 0.13  & 2.61  \\
        Prompt Injection        & 1.81  & 0.96  & 0.54  & 0.91  \\
        Social Engineering      & 0.00  & 0.64  & 0.67  & 0.55  \\
        Resource Abuse          & 0.00  & 0.00  & 0.13  & 0.06  \\

        \midrule
        \rowcolor{oursbg}
        \textbf{Total}
        & \textbf{100.00}
        & \textbf{100.00}
        & \textbf{100.00}
        & \textbf{100.00} \\

        \bottomrule
    \end{tabular}
\end{table}
Table~\ref{tab:category-risk-distribution} summarizes the threat categories identified by SkillScan during the Phase~1 cold-start evolution. SkillScan assigns each detected finding both a threat category and a risk level, where the risk indicates the relative risk associated with that finding. Accordingly, the same threat category may appear under different risk levels depending on the specific behavior detected. In this analysis, we retain the High, Medium, and Low risk labels reported by SkillScan and refer to them as risk levels. All category and risk labels are taken directly from the original SkillScan outputs, without manual relabeling or post-hoc normalization.

The results show that the findings observed during cold-start evolution span a broad range of threat categories rather than concentrating on a single attack pattern. Overall, Data Exfiltration and Unauthorized Tool Use are the two most frequent categories, accounting for 23.65\% and 23.59\% of all findings, respectively, followed by Skill Discovery Abuse at 17.53\%. The category composition also varies across risk levels. High-risk findings are concentrated in Data Exfiltration (26.81\%) and Command Injection (23.55\%), with Policy Violation accounting for another 14.49\%. Medium-risk findings are primarily associated with Unauthorized Tool Use (30.14\%) and Data Exfiltration (28.87\%), whereas low-risk findings are more broadly distributed across Unauthorized Tool Use (23.06\%), Skill Discovery Abuse (22.65\%), Data Exfiltration (18.10\%), and Harmful Content (16.22\%). These statistics characterize the security signals surfaced by SkillScan during the initial scanner-guided cold-start process and provide additional context for the subsequent evolution.

% Required packages: \usepackage{booktabs,wrapfig}

\section{Harness Comparison}
\label{app:harness-comparison}

\begin{wraptable}{r}{0.35\textwidth}
    \centering
    \setlength{\tabcolsep}{5pt}
    \renewcommand{\arraystretch}{1.1}
    \caption{
        Harness comparison.
        All values are percentages.
    }
    \label{tab:harness-comparison}
    \begin{tabular}{@{}lcc@{}}
        \toprule
        \textbf{Harness}
        & \textbf{ASR} $\uparrow$
        & \textbf{DR} $\downarrow$ \\
        \midrule
        Codex   & 46.99 & 0.00 \\
        Claude Code & 42.57 & 0.00 \\
        \bottomrule
    \end{tabular}
\end{wraptable}

To examine the sensitivity of SkillDRE to the execution harness,
we compare Codex~\citep{openaiCodexCLI} and
Claude Code~\citep{anthropicClaudeCode}, using DeepSeek-V4-Pro
as the victim model in both settings.
Table~\ref{tab:harness-comparison} reports skill-level ASR and DR,
following the definitions in the main experiments.
DR measures detection by SkillScan on the final submitted
candidate Skills, rather than runtime detection by either harness.

SkillDRE achieves an ASR of 46.99\% with Codex and 42.57\% with Claude Code.
ASR is 4.42 percentage points lower with Claude Code, while SkillScan DR remains 0\% in both settings.
These results show that SkillDRE can realize malicious objectives under both tested harnesses while retaining scanner acceptance, indicating that its effectiveness is not confined to Codex.
\end{document}